\documentclass[letterpaper,11pt]{article}
\usepackage{preamble}

\newcommand{\lambdaonep}{\ensuremath{\lambda_1^{(p)}}}

\begin{document}
\hypersetup{pageanchor=false}

\title{How to Round Spiky Balls into Submission \\
\large{And Faster Algorithms For The Closest Vector Problem}
}
	\author{Divesh Aggarwal\\
		National University of Singapore\\
		\texttt{dcsdiva@nus.edu.sg}
        \and
        Aditya Morolia\\
        Centre for Quantum Technologies, Singapore\\
        \texttt{morolia@u.nus.edu}
			 }
	\date{\today}
	\maketitle

\begin{abstract}
    We give the first single-exponential time algorithm for the exact Closest Vector Problem (CVP) in the $\ell_p$ norm for $1 < p < 2$. 
    Prior to this, such an algorithm was only known for the Euclidean norm, i.e., $\ell_2$, or needed an additional promise that the target is close to the lattice. 
    Using similar techniques, we also find a shortest non-zero vector in the $\ell_p$ norm in time $2^{O(n)}$ for every $p \in [1,2)$ with a constant in the exponent that doesn't depend on $p$, and that improves state of the art algorithms for the shortest vector problem in the $\ell_p$ norm.

    The problem of finding a single exponential time algorithm for CVP in the $\ell_p$ norm for $p \neq 2$ has been open since 2010, when Micciancio and Voulgaris resolved it for the Euclidean norm.

    One of our key ideas is that for $0<p\le2$, the function $z \mapsto e^{-|z|^p}$ is a mixture of Gaussians, so the $\ell_p$ discrete Gaussian distribution on a lattice shift is a mixture of Euclidean discrete Gaussians with random diagonal widths.
    Our algorithms use this to reduce the $\ell_p$ problem to the Euclidean CVP on a randomly scaled lattice. 
\end{abstract}

\newpage
{\setlength{\parskip}{2pt}
    \tableofcontents}
	\setcounter{page}{0}
	\thispagestyle{empty}
	\newpage
	\setcounter{page}{1}
    \hypersetup{pageanchor=true}
	
	\section{Introduction}
	\label{sec:intro}
    	A lattice $\cL\subseteq\R^n$ is the set of all integer linear combinations of $k$ linearly independent vectors $\vec b_1,\dots,\vec b_k \in \R^n$.
We write 
\[ \cL(B) \coloneqq \set{ B\vec z : \vec z \in \Z^k } \]
for the lattice generated by the basis $B$, which is a matrix that has $\vec b_i$ as the $i$th column.
We say $k$ is the rank of $\cL$, $n$ its dimension, and we say that $\cL$ has full rank if $k=n$. 
Given a basis and a norm, the \emph{Shortest Vector Problem}
($\svp$) asks for a shortest nonzero vector in $\cL$, and the \emph{Closest
Vector Problem} ($\cvp$) asks, given in addition a target $\vec t\in\R^n$, for
a vector in $\cL$ closest to $\vec t$. 
In this work, we study algorithms for them in the $\ell_p$ norm. For $1 \leq p \leq \infty$, and $\vec x \in \R^n$, the $\ell_p$ norm of $\vec x$ is 
$$\norm{\vec x}_p=\paren{\sum_j|x_j|^p}^{1/p},$$
and we write $\svp_p$ and $\cvp_p$ when the length of the vectors is measured in the $\ell_p$ norm.

These problems are fundamental in the algorithmic
geometry of numbers \cite{MG02,RegevCourse,Eisenbrand10,SD23complexity}.
Algorithms for them, exact and approximate, are used in polynomial factorization
and basis reduction \cite{LLL82,Babai86,Schnorr87,GN08,ALNS20,AEPS25} and closest-point decoding in communication systems \cite{AEVZ02}. 
Integer programming, which is a discrete optimization problem of particular importance, can be studied as a lattice problem \cite{DBLP:journals/mor/Lenstra83,DBLP:conf/stoc/Kannan83,Kannan1987MinkowskisCB,Frank1987AnAO,DadushThesis12,DBLP:conf/focs/ReisR23}.
In the $\ell_2$ norm, they are extremely important in modern cryptography.
Their conjectured hardness is the foundation of lattice-based cryptography, through
the worst-case to average-case reductions of Ajtai and of Regev and the follow-up work
\cite{Ajtai96,AD97,Micciancio04,MR07,Regev09,Peikert09,GPV08,MP12,BLPRS13,LPR13};
see the surveys \cite{MR09,Peikert16}. 
Lattice-based key
encapsulation is now a NIST standard \cite{FIPS203}, and the running time of
the best exact algorithms is what determines their concrete parameters.

\paragraph{Hardness Results.}
Exact $\cvp$ is NP-hard in every $\ell_p$ norm \cite{vEB81}; approximation is
hard within constant factors \cite{ABSS97} and within $n^{c/\log\log n}$
\cite{DKRS03}, and remains hard with preprocessing
\cite{Micciancio01cvpp,FM04,Regev04TIT,AKKV11,KPV12}; for $\svp$ see
\cite{Ajtai98,CN99,Micciancio01,Khot05,HR12,Micciancio12,BP23} and, for
$\ell_\infty$, \cite{Dinur02}. A recent industrial manuscript announces a
deterministic reduction giving $n^{1/(200p)}$ hardness for each fixed rational
$p\ge1$ \cite{OpenAI26}. In the other direction, Euclidean $\gcvp$ is in
coAM at factor $O(\sqrt{n/\log n})$ \cite{GG00} and in $\np\cap$co$\np$ at
factor $O(\sqrt n)$ \cite{AR05}, and Peikert extended limits of this kind to
$\ell_p$ norms \cite{Peikert08}. Fine-grained lower bounds make the role of
the dimension explicit. Exact $\cvp_p$ requires $2^{(1-\rho)n}$ time under
SETH for every $\rho>0$ and every finite $p\notin2\Z$, with a constant-factor
counterpart under Gap-SETH \cite{BGS17,ABGS21}; constant-factor $\cvp_p$
requires $2^{\Omega(n)}$ time under ETH for every finite $p$ \cite{AGMZ26},
see also \cite{HKW26}. For $\svp$, Aggarwal and Stephens-Davidowitz showed
that $\svp_p$ has no $2^{o(n)}$-time algorithm for $p>2$ under Gap-ETH and
gave SETH-based bounds for large $p$, while for $1\le p\le2$ their lower
bounds need an additional assumption on lattices with exponential kissing
number \cite{AS18stoc}; see \cite{BPT22,AGMZ26} for later improvements and
\cite{AC21,ABBGKLPSV23} for related fine-grained results. For $\svp_p$ with
$1\le p\le2$ the evidence for an exponential lower bound is therefore
conditional.


\paragraph{Algorithms in the Euclidean norm.}
These problems are most well studied for the Euclidean ($\ell_2$) norm. The polynomial-time basis reduction algorithm of \cite{LLL82} approximates both problems to within exponential, or slightly subexponential,
factors (see \cite{Babai86,Schnorr87,GN08,ALNS20,AEPS25}).
Algorithms for the exact versions of these problems can be categorized in four families. 
Enumeration, due to Pohst, Kannan, and
Fincke
\cite{Poh81,DBLP:conf/stoc/Kannan83,FP85,Kannan1987MinkowskisCB}, with
refinements by Helfrich \cite{Hel85} and the sharp analysis of Hanrot and
Stehl\'e \cite{HS07}, solves $\svp$ in time $n^{n/(2e)+o(n)}$ and $\cvp$ in
time $n^{n/2+o(n)}$ using polynomial space; Micciancio and Walter have a more refined analysis \cite{MW15}. 
Lattice sieving, introduced by Ajtai, Kumar and
Sivakumar \cite{AKS01,AKS02}, gave the first randomized $2^{O(n)}$-time
algorithm for $\svp$ and for approximate $\cvp$; its (provably correct) constant in the exponent was
analyzed and improved in \cite{NV08,MV10sieve,PS09}. Micciancio and
Voulgaris \cite{MV13} gave a deterministic $2^{2n+o(n)}$-time algorithm for
exact $\cvp$ and hence for most other lattice problems, by computing the Voronoi cell;
see \cite{BD15,HRS20} for further work on Voronoi cells. Aggarwal, Dadush,
Regev and Stephens-Davidowitz \cite{ADRS} solved $\svp$ in time $2^{n+o(n)}$
by sampling from the discrete Gaussian distribution, and Aggarwal, Dadush and
Stephens-Davidowitz \cite{ADS} achieved the same time complexity for exact $\cvp$ in the Euclidean norm; Aggarwal and
Stephens-Davidowitz simplified both \cite{AS18sosa}. Recent work gives further improvements for Euclidean $\svp$ and discrete
Gaussian sampling \cite{GFH26,GFH26b,Hhan26,Hhan26b,Kim26}. Our algorithms
use the exact Euclidean algorithms of \cite{ADRS,ADS,MV13} as black boxes.

\paragraph{Algorithms for other $\ell_p$ norms.}
It is perhaps not a coincidence that historically it has been much easier to discover algorithms for the Euclidean norm. The $\ell_2$ norm embeds into the $\ell_p$ norm with
distortion arbitrarily close to $1$, so $\cvp_2$ and $\svp_2$ reduce to their
$\ell_p$ analogues at the cost of a factor $1+\varepsilon$ \cite{RR06} in the approximation factor; and
the fine-grained lower bounds above are strongest exactly for $p\notin2\Z$
\cite{BGS17,ABGS21}, a range that contains the interval $1<p<2$ studied here.
The $\ell_1$ norm is of particular interest because integer programming
reduces to exact $\cvp_1$ (see \Cref{sec:related-work}).

For $\svp$ the single-exponential barrier was broken long ago. The sieve of
Ajtai, Kumar and Sivakumar extends to every $\ell_p$ norm, as shown by
Bl\"omer and Naewe \cite{BN09} and by Arvind and Joglekar \cite{AJ08}, with
improved constants given by Mukhopadhyay \cite{Muk21} and, for
$\ell_\infty$, in \cite{EHN11,AM18}. Dadush, Peikert and
Vempala \cite{dadush2011enumerativelatticealgorithmsnorm} gave
$2^{O(n)}$-time algorithms for exact $\svp$ in every norm, by enumerating
lattice points with coverings by an $M$-ellipsoid of the unit ball, which
they compute in expected $2^{O(n)}$ time for any convex body and in
deterministic polynomial time for $\ell_p$ balls; Dadush and Vempala made the
general construction deterministic \cite{DV12,DV13,DadushThesis12}.
Constant-factor approximations of $\svp$ and $\cvp$ take time
$2^{(0.802+\xi)n}$ in $\ell_p$ norms \cite{EV22} and in every norm
\cite{RV22}.

For $\cvp$ the picture is different: every single-exponential algorithm for
a norm other than $\ell_2$ is either approximate or needs a promise on the
target. The method of Dadush, Peikert and Vempala solves exact $\cvp$ in
single-exponential time only when the target lies within a constant multiple
$\gamma$ of $\lambda_1(\cL)$ of the lattice, at a cost that grows with
$\gamma$ \cite{dadush2011enumerativelatticealgorithmsnorm,DV12}; such
promise versions are closely related to bounded distance decoding
\cite{LLM06,DRS14}. Arvind and Joglekar obtain a promise result of this kind
by sieving, and show that unrestricted exact $\cvp$ would follow from a
version in which lattice vectors attaining the successive minima are given
\cite{AJ08}. Approximation algorithms are known and there is a tradeoff between the approximation factor and the time complexity (see \Cref{tab:approx}). Constant approximation factors are achieved in time
$2^{O(n)}$ \cite{BN09,DK16} and $2^{(0.802+\xi)n}$ \cite{EV22,RV22}. For
$(1+\varepsilon)$-approximation the running times are
$(2+1/\varepsilon)^{O(n)}$ for $\ell_p$ norms by sieving \cite{BN09};
$2^{O(n)}(1+1/\varepsilon)^n$, deterministically and in every norm, by
lattice sparsification \cite{DK16}; $2^{O(n)}\varepsilon^{-n/\min\{2,p\}}$
for $\ell_p$ norms by covering \cite{NV22}; and
$2^{O(n)}(\log(1/\varepsilon))^{O(n)}$ for $\ell_\infty$ \cite{EHN11} and for
polyhedral norms with $O(n)$ facets \cite{NV22}. In every case the exponent
of the $\varepsilon$-dependence grows linearly with the dimension. 

Exact algorithms for $\cvp$ in non-Euclidean norms are
superexponential. 
To see this, consider a lattice basis $B$ and a target $\vec t$. 
For a threshold $R > 0$, consider the convex body 
\[
 K_R \coloneqq \set{\vec z\in\R^n:\norm{B\vec z-\vec t}_p\le R}.
\]
$K_R$ contains an integer point if and only if there exists a lattice vector within a distance $R$ from $\vec t$ (in the $\ell_p$ norm). 
This is the well known Integer Programming problem with $n$ variables. 
Lenstra's algorithm \cite{DBLP:journals/mor/Lenstra83}, with
Kannan's improvement \cite{DBLP:conf/stoc/Kannan83,Kannan1987MinkowskisCB}
and later refinements \cite{Frank1987AnAO,HK13,DadushThesis12,DER25}, gives a
$n^{O(n)}$ time algorithm for this problem.
The recent work of Reis and Rothvoss
\cite{DBLP:conf/focs/ReisR23} gives a $(\log 2n)^{O(n)}$ time algorithm, which is currently the best known.
Bl\"omer and Naewe \cite{BN11} gave a deterministic, polynomial-space
algorithm for exact $\cvp_p$ in every $\ell_p$ norm with $1<p<\infty$, by generalizing Lenstra's algorithm to a lattice
membership problem; for $n$-dimensional lattices and targets in their span it
runs in time $n^{(2+o(1))n}$ times a polynomial.

Whether exact $\cvp$ in some norm other than $\ell_2$ can be solved in
single-exponential time has therefore remained open. Micciancio and Voulgaris
asked whether their techniques extend beyond the Euclidean norm \cite{MV13};
Dadush, Peikert and Vempala, asked whether their enumeration approach extends beyond the promise
\cite{dadush2011enumerativelatticealgorithmsnorm,DV12}.
Bl\"omer and
Kohn \cite{BK18} call the questions of a single-exponential exact $\cvp$ in a norm other than $\ell_2$
``the main open question in this area''. They showed that the Voronoi-cell
approach cannot be used for other $\ell_p$ norms as it stands, because already in $\R^3$ the
number of $\ell_3$-Voronoi-relevant vectors of a lattice is unbounded
\cite{BK18}. The other Euclidean methods do not transfer either. 
One could also try to extend the discrete Gaussian sampling approach of \cite{ADS,AS18sosa}.
The $\ell_p$
analogue of the discrete Gaussian, with the probability of a vector $\vec x \in \cL - \vec t$ proportional to
$e^{-\norm{\vec x}_p^p / r^p}$, would reveal closest vectors when sampled at a small enough
parameter $r$, and indeed sampling from it is \emph{equivalent} to $\cvp_p$
\cite[Section~5]{SD16}, so one cannot sample from it first and then execute ideas similar to those for algorithms in the Euclidean norm.
Direct reductions between norms \cite{RR06,ACKLS21} lose
approximation factors and cannot produce exact solutions.

\paragraph{This paper.}
We obtain the first single-exponential
algorithm for exact $\cvp$ in a non-Euclidean $\ell_p$ norm for $1<p<2$ without a promise
on the target. 
We do so by establishing a new connection between the discrete Gaussian distribution and a ``discrete super-Gaussian" distribution over lattice shifts.
The same technique gives a short algorithm for exact $\svp_p$
for every $p$ satisfying $1\le p<2$, with an improved constant in the exponent compared to prior work.

\subsection{Our results}
\label{sec:results}

We write $|I|$ for the bit length of the input, $\dist_p(\cL,\vec t)$ for the
$\ell_p$ distance of $\vec t$ to $\cL$, and $\lambdaonep(\cL)$ for the
$\ell_p$ length of a shortest nonzero vector of $\cL$.  

\begin{theorem}[Exact $\cvp_p$]
\label{thm:main-cvp}
Fix $p\in(1,2)$. There is a randomized algorithm that, given a basis of a
lattice $\cL\subseteq\Q^n$ of rank $n$, a target $\vec t\in\Q^n$ and
$\delta\in(0,1/2]$, outputs a vector of $\cL$ closest to $\vec t$ in the
$\ell_p$ norm with probability at least $1-\delta$, in expected time
$2^{O_p(n)}\cdot\poly(|I|)\cdot\log(1/\delta)$.
\end{theorem}

\Cref{thm:main-cvp} is proved in \Cref{sec:4_cvp} (\Cref{thm:cvp}). Together
with the Euclidean case \cite{MV13,ADS}, it shows that exact $\cvp_p$ can be
solved in single-exponential time for every $p\in(1,2]$. As far as we know,
it is the first algorithm for exact $\cvp$ in a non-Euclidean $\ell_p$ norm
that runs in single-exponential time\footnote{We note that our constants in the exponent for both $\cvp$ and $\svp$ can in fact be improved. We chose the current proofs as the best exposition of our new techniques.} without any promise on the distance of the target. The previous bounds for exact $\cvp_p$ with $1<p<2$ were
$n^{(2+o(1))n}$ for targets in the span \cite{BN11} and, through convex
integer programming \cite{DBLP:conf/focs/ReisR23} combined with a standard
root-separation bound, $(\log 2n)^{O(n)}\cdot2^{O_p(n)}$. 

\begin{theorem}[Exact $\svp_p$]
\label{thm:main-svp}
For every $p\in[1,2)$ there is a randomized algorithm that, given a basis of
a lattice $\cL\subseteq\Q^n$ of rank $n$, outputs a shortest nonzero vector of
$\cL$ in the $\ell_p$ norm with probability at least $1-e^{-n}$, in time at most $2^{2.4368 n + o(n)}\cdot\poly(|I|)$.
\end{theorem}

For $p=2$ the corresponding statement is the theorem of \cite{ADRS}, so exact
$\svp_p$ is single-exponential for every $1\le p\le2$. 

We assume that the input is a full rank lattice only for simplicity. For a lattice of rank $k<n$ in
$\Q^n$ the same $\svp$ algorithm runs in time $2^{O_p(k)}\poly(|I|)$, while the $\cvp$ algorithm runs in time $2^{O_p(n)}\poly(|I|)$.

\begin{table}[H]
\centering
\small
\renewcommand{\arraystretch}{1.15}
\begin{tabular}{p{0.25\textwidth}p{0.40\textwidth}p{0.25\textwidth}}
\hline
\raggedright Work & \raggedright Problem and setting & \raggedright Exponential factor \tabularnewline
\hline
\raggedright Kannan; Hanrot--Stehl\'e \cite{Kannan1987MinkowskisCB,HS07} & \raggedright exact $\svp_2$, $\cvp_2$; deterministic, polynomial space & \raggedright $n^{n/(2e)+o(n)}$, $n^{n/2+o(n)}$ \tabularnewline
\raggedright Micciancio--Voulgaris \cite{MV13} & \raggedright exact $\cvp_2$, $\svp_2$; deterministic & \raggedright $2^{2n+o(n)}$ \tabularnewline
\raggedright Aggarwal, Dadush, (Regev,) Stephens-Davidowitz \cite{ADRS,ADS} & \raggedright exact $\svp_2$, $\cvp_2$ & \raggedright $2^{n+o(n)}$ \tabularnewline
\raggedright Bl\"omer--Naewe; Arvind--Joglekar; Mukhopadhyay \cite{BN09,AJ08,Muk21} & \raggedright exact $\svp_p$ by sieving & \raggedright $2^{O(n)}$; $2^{2.751n+o(n)}$ in \cite{Muk21} \tabularnewline
\raggedright Dadush--Peikert--Vempala \cite{dadush2011enumerativelatticealgorithmsnorm,DV12,DV13} & \raggedright exact $\svp$ in any norm; exact $\cvp$ if $\dist(\vec t,\cL)\le\gamma\lambda_1(\cL)$ & \raggedright $2^{O(n)}$; growing with $\gamma$ \tabularnewline
\raggedright Bl\"omer--Naewe \cite{BN11} & \raggedright exact $\cvp_p$, $1<p<\infty$; deterministic; $\vec t\in\operatorname{span}(\cL)$ & \raggedright $n^{(2+o(1))n}$ \tabularnewline
\raggedright Reis--Rothvoss \cite{DBLP:conf/focs/ReisR23} & \raggedright convex integer feasibility in $n$ variables; with root separation, exact $\cvp_p$ & \raggedright $(\log2n)^{O(n)}$ \tabularnewline
\raggedright \textbf{This paper} & \raggedright exact $\cvp_p$, $1<p<2$ & \raggedright $2^{O_p(n)}$ \tabularnewline
\raggedright \textbf{This paper} & \raggedright exact $\svp_p$, $1\le p<2$ & \raggedright $2^{O(n)}$, constant uniform in $p$ \tabularnewline
\hline
\end{tabular}
\caption{Exact algorithms in dimension $n$. Polynomial factors in the input length are suppressed.}
\label{tab:exact}
\end{table}

\begin{table}[H]
\centering
\small
\renewcommand{\arraystretch}{1.15}
\begin{tabular}{p{0.25\textwidth}p{0.40\textwidth}p{0.25\textwidth}}
\hline
\raggedright Work & \raggedright Problem and setting & \raggedright Exponential and accuracy dependence \tabularnewline
\hline
\raggedright Bl\"omer--Naewe \cite{BN09} & \raggedright $(1+\varepsilon)$-$\cvp$, $\ell_p$ norms & \raggedright $(2+1/\varepsilon)^{O(n)}$ \tabularnewline
\raggedright Dadush--Kun \cite{DK16} & \raggedright $(1+\varepsilon)$-$\cvp$, any norm; deterministic & \raggedright $2^{O(n)}(1+1/\varepsilon)^{n}$ \tabularnewline
\raggedright Nasz\'odi--Venzin \cite{NV22} & \raggedright $(1+\varepsilon)$-$\cvp$, $\ell_p$ norms & \raggedright $2^{O(n)}\varepsilon^{-n/\min\set{2,p}}$ \tabularnewline
\raggedright Arya--da Fonseca--Mount \cite{ADM24} & \raggedright $(1+\varepsilon)$-$\cvp$, well-centered convex gauges & \raggedright $2^{O(n)}\varepsilon^{-(n-1)/2}$ \tabularnewline
\raggedright Eisenbrand--H\"ahnle--Niemeier \cite{EHN11}; Nasz\'odi--Venzin \cite{NV22} & \raggedright $(1+\varepsilon)$-$\cvp$, $\ell_\infty$; polyhedral norms with $O(n)$ facets & \raggedright $2^{O(n)}(\log(1/\varepsilon))^{O(n)}$ \tabularnewline
\raggedright Eisenbrand--Venzin \cite{EV22} & \raggedright $O_\xi(1)$-$\cvp$, $\ell_p$ norms & \raggedright $2^{(0.802+\xi)n}$ \tabularnewline
\raggedright Rothvoss--Venzin \cite{RV22} & \raggedright $O_\xi(1)$-$\cvp$ and $\svp$, any norm; rank $k$ in arbitrary dimension & \raggedright $2^{(0.802+\xi)k}$ \tabularnewline
\hline
\end{tabular}
\caption{Approximation algorithms for $\cvp$ in non-Euclidean norms, compared by their exponential factor and their dependence on the accuracy.}
\label{tab:approx}
\end{table}

\paragraph{The range of $p$.}
For $0<p\le2$ the function $e^{-|s|^p}$ is a mixture of centered Gaussians,
with the mixing defined by a positive $(p/2)$-stable distribution
\cite{Feller71,west_scale_mixtures,nolan2020}; this is what lets us reduce to Euclidean
problems in a random ``diagonal metric". For $p>2$, no such representation exists
\cite{Schoenberg_psd}. The $\cvp$ algorithm additionally needs $p>1$: the step that
transfers the analysis from the unknown closest vector to the vector the
algorithm currently holds uses the uniform convexity of $s\mapsto|s|^p$, and
its constant grows boundlessly as $p\to1$.

\subsection{Technical Overview}
\label{sec:techniques}

Let $1 \leq p < 2$.
Given a full rank lattice $\cL \subset \R^n$ and a target $\vec t\in \R^n$, our goal is to find a \[ \vec v^* \in \argmin_{\vec x \in \cL - \vec t} \norm{\vec x}_p^p . \]

We already know how to solve the $\cvp$ problem in the Euclidean metric ($\ell_2$) in single exponential time \cite{MV13,ADS}. 
A key subroutine in the algorithm of \cite{ADS} is an (approximate) sampler for a probability distribution $D_2(\cL - \vec t, r)$ over $\cL - \vec t$ such that 
 \[ \Pr_{X \sim D_2(\cL - \vec t, r)}[X = \vec x] \propto \rho(\vec x, r) \coloneqq \exp\paren{- \pi \frac{\norm{x}_2^2}{r^2}}. \]


Therefore, a natural approach is to consider analogous distributions in the $\ell_p$ metric: $D_p(\cL - \vec t, r)$ over $\cL - \vec t$ such that 
 \[ \Pr_{X \sim D_p(\cL - \vec t, r)}[X = \vec x] \propto f_p(\vec x, r) \coloneqq \exp\paren{-\frac{\norm{x}_p^p}{r^p}}, \] 
and sample from them for a small enough $r$. This family of distributions is known as a super-Gaussian, and has been previously considered in reductions between lattice problems \cite{SDthesis17,ACKLS21}. 

\begin{figure}[!t]
    \centering
    \begin{minipage}[t]{0.49\linewidth}
        \centering
        \includegraphics[width=\linewidth]{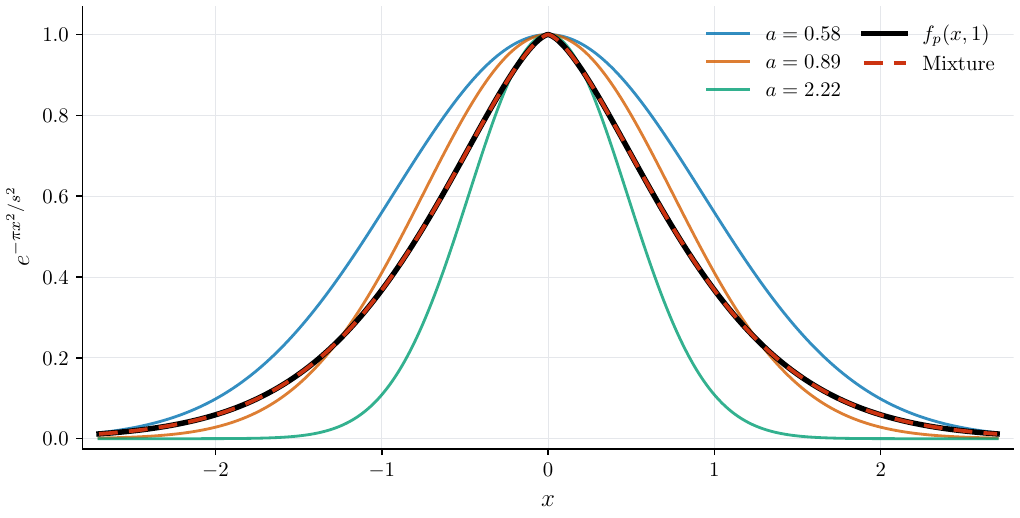}
        \par\smallskip (a)
    \end{minipage}\hfill
    \begin{minipage}[t]{0.49\linewidth}
        \centering
        \includegraphics[width=\linewidth]{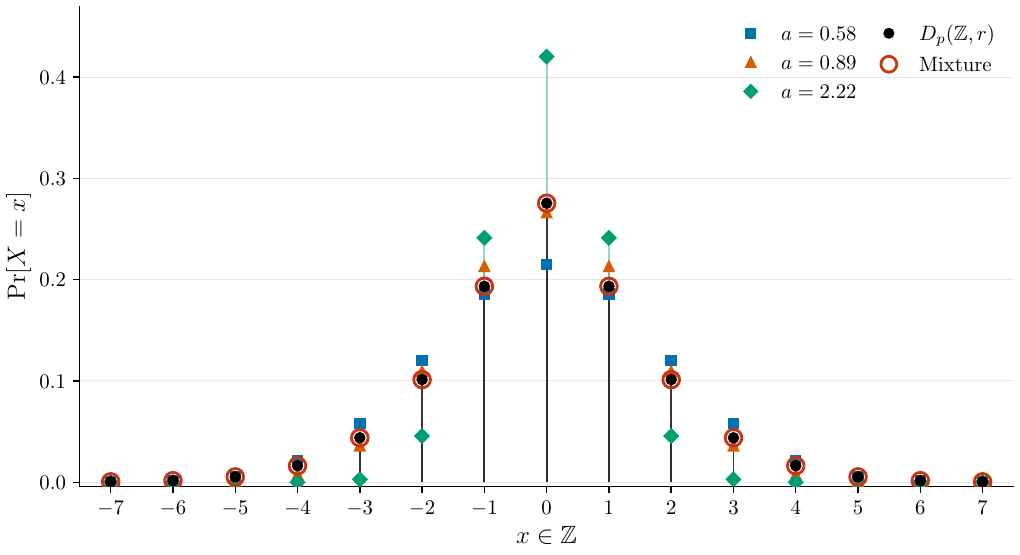}
        \par\smallskip (b)
    \end{minipage}
    \caption[Gaussian scale mixture for the super-Gaussian]{\textbf{Approximating a super-Gaussian with a variance mixture of Gaussians.}
    Here $p=1.5$ and $a=\pi r^2/s^2$. \textbf{(a)} For $r=1$, the black curve is $f_p(x,1)=e^{-|x|^p}$;
    the thin colored curves are Gaussian functions $e^{-ax^2}$,
    with $a$ as indicated in the legend.
    The dashed red curve averages $N=120{,}000$ such functions with independent
    $a\sim g_{p/2}$, illustrating the identity
    $f_p(x,1)=\int_0^\infty e^{-\pi x^2/s^2}h_p(s,1)\,ds$.
    \textbf{(b)} On $\Z$ with $r=2$, normalized discrete Gaussians mix to $D_p(\Z,r)$ using scale weights proportional to $h_p(s,r)\rho(\Z,s)$.}
    \label{fig:gaussian_scale_mixture}
\end{figure}

To this end, our starting point is the following identity. For $x \in \R$ and $r > 0$, we have that 

\begin{equation}
\label{eqn:mixture_identity_intro}
    \exp\paren{-\frac{\abs{x}^{p}}{r^p}}
=
\int_0^\infty \exp\paren{-\pi \frac{x^2}{s^2}} g_{p/2}\paren{\frac{\pi r^2}{s^2}} \frac{2\pi r^2}{s^3} ds,
\end{equation}
where $g_{p/2}(\cdot)$ is the density of the $p/2$-stable distribution \cite{AndrewsMallows74,west_scale_mixtures,nolan2020}. 
See \cref{ssec:stabel_dist} for a short proof of this identity and \cref{fig:gaussian_scale_mixture} for an illustration.
Using this identity coordinate-wise, super-Gaussian distributions can be written as a marginal distribution of a {\em variance mixture} of the Gaussian distribution:

\[ f_p(\vec x, r) = \int_{\R^n_{>0}} \rho(\vec x, \vec s) h_p(\vec s, r) d\vec s, \]
where \[ \rho(\vec x, \vec s) \coloneqq \exp\paren{-\pi  \sum_j \frac{x_j^2}{s_j^2}}; \qquad h_p(\vec s, r) \coloneqq \prod_{j=1}^n
g_{p/2}\!\left(\frac{\pi r^2}{s_j^2}\right)
\frac{2\pi r^2}{s_j^3}. \]

Using this result, sampling from a joint distribution over $(\cL - \vec t) \times \R_{> 0}^n$ already allows us to solve exact $\svp_p$ and constant approximate $\cvp_p$ for $p \in [1, 2)$ in $2^{O(n)}$ time. 
However, we were unable to extend this approach to exact \cvp.
Define \[ F_p(\vec x) \coloneqq \sum_j \abs{x_j}^p. \]
A geometric barrier we face is the fact that in the $\ell_2$ metric, the Hessian is \[
\nabla^2F_2(\vec x)=2I,
\]
whereas for \(1<p<2\), for non-zero coordinates, 
\[
\nabla^2F_p(\vec x)
=p(p-1)\operatorname{diag}(|x_j|^{p-2}).
\]
Thus the curvature of the objective function depends on where we evaluate it. Crucially, it misbehaves as we move closer to the ``spiky" corners.

\begin{figure}[!t]
    \centering
    \begin{minipage}[t]{0.49\linewidth}
        \centering
        \includegraphics[width=\linewidth]{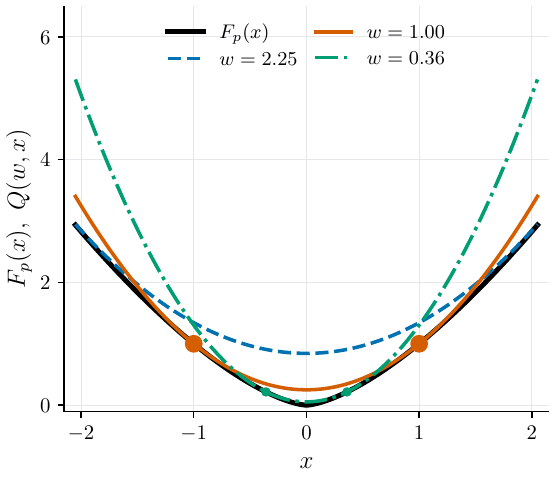}
        \par\smallskip (a)
    \end{minipage}\hfill
    \begin{minipage}[t]{0.49\linewidth}
        \centering
        \includegraphics[width=\linewidth]{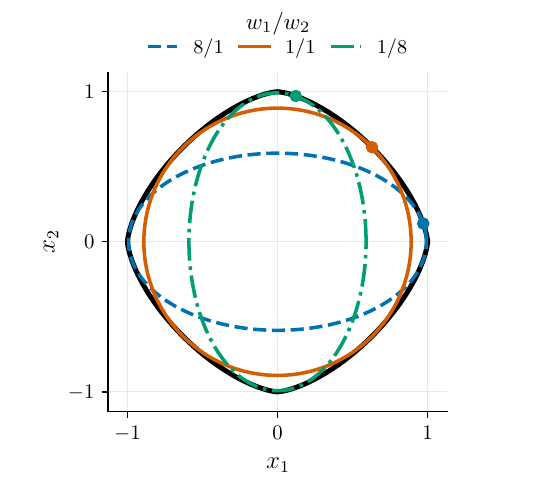}
        \par\smallskip (b)
    \end{minipage}
    \caption[Quadratic upper bounds and their weighted Euclidean geometry]{\textbf{Approximating the $\ell_{1.5}$ objective with weighted $\ell_2$ objectives.}
    We consider $p=1.5$.
    \textbf{(a)} The black curve is $F_p(x)=|x|^p$; the colored curves are
    $Q(w,x)=\frac p2|w|^{p-2}x^2+(1-\frac p2)|w|^p$ for various values of $w$ as in the legend.
    Each quadratic lies above $F_p$ and touches it at $x=\pm w$.
    \textbf{(b)} The black curve bounds the unit $\ell_p$ ball.
    The colored ellipses are $Q(\vec w,\vec x)=1$, with $F_p(\vec w)=1$
    and the ratios $w_1 / w_2$ as indicated.
    Since $F_p(\vec x)\le Q(\vec w,\vec x)$, each ellipsoid lies inside the
    $\ell_p$ ball and is tangent at every coordinatewise reflection of $\vec w$.
    Its boundary is $\sum_{j=1}^2|w_j|^{p-2}x_j^2=1$.
    Equivalently, setting $s_j^2=2\pi|w_j|^{2-p}/p$ (with $r=1$) gives
    $\pi\sum_{j=1}^2x_j^2/s_j^2=p/2$: varying the scales changes the Euclidean geometry.}
    \label{fig:quadratic_geometry}
\end{figure}

\textbf{Approximating the $\ell_p$ geometry by $\ell_2$.}
We will show that \cref{eqn:mixture_identity_intro} can in fact be used to define a natural distribution over weighted $\ell_2$ balls that approximate a target $\ell_p$ ball. 
Therefore, one can replace the $\ell_p$ objective function $F_p(\cdot)$
with {\em a mixture of weighted $\ell_2$} objective functions. See \cref{fig:quadratic_geometry}.
A similar approach has been previously considered\footnote{A major role of AI tools in this work was in fact to direct us to this beautiful line of work.}, for example, in the Iteratively Weighted Least Squares literature, see \cite[Section 5.7.1]{Gentle2024}.

For any $\vec w, \vec x \in \R^n$ such that $\forall j,~w_j \neq 0$, let us define 
\[ Q(\vec w, \vec x) \coloneqq \frac{p}{2} \sum_j \abs{w_j}^p \paren{\frac{x_j}{w_j}}^2 + \paren{1 - \frac{p}{2}} F_p(\vec w). \]
By concavity of $z\mapsto z^{p/2},~ z \geq 0$, we have \[ F_p(\vec x) \leq Q(\vec w, \vec x), \]
with equality if and only if $\forall j : \abs{w_j} = \abs{x_j}$. 
The important observation is that, for any fixed $\vec w$, $Q(\vec w,\cdot)$ is a quadratic function. Therefore, minimizing $Q(\vec w,\cdot)$ over a lattice shift is simply an instance of the Euclidean $\cvp_2$, which we can solve in $2^{n}$ time using \cite{ADS}.

Given a $\vec x \in \cL - \vec t$, the Gaussian mixture identity naturally defines a posterior over the weights, proportional to 
\[
P_{\vec x,r}(\vec s)
\coloneqq
\frac{\rho(\vec x,\vec s)\,h_p(\vec s,r)}
     {f_p(\vec x,r)}.
\]
Let \(\vec s\) be a sample from this distribution, and set
\[
w_j=\left(\frac{p s_j^2}{2\pi r^p}\right)^{1/(2-p)}.
\]

The key property of this choice is that for $$ E(\vec v^*, \vec w) \coloneqq Q(\vec w,\vec v^*)-F_p(\vec v^*), $$
we have 
\[
\mathbb E_{\vec s\sim P_{\vec v^*,r}}
   \bigl[E(\vec v^*,\vec w)\bigr]
\le 2\ln 2\,nr^p.
\]
If $\vec y$ minimizes $Q(\vec w,\cdot)$ over the coset containing $\vec v^*$, then
$$ F_p(\vec y) \le Q(\vec w,\vec y) \le Q(\vec w,\vec v^*) = F_p(\vec v^*) + E(\vec v^*, \vec w). $$
Thus, 
\[
\mathbb E_{\vec s\sim P_{\vec v^*,r}}\bigl[F_p(\vec y)-F_p(\vec v^*)\bigr]
\le 2\ln 2\,nr^p.
\]


We show a similar result in \cref{lem:closest_is_best}.
Suppose for a moment we had a way to sample ``good quadratic objectives'' $Q(\vec w, \cdot)$ from a distribution close to $P_{\vec v^*, r}$ using an approximate shortest vector $\vec x \in 2\cL + \vec v^*$ as a hint. 
Then, we could follow an iterative approach: 
starting from $\vec x$, we iteratively find a successor $\vec y$ that minimizes $Q(\vec w, \cdot)$ for many choices of $\vec w$, and keep the shortest. 
Finally, we can (relatively easily) show that once $\vec y$ is short enough (in the correct coset), we are done. 
Of course, we don't know which coset $\vec v^*$ lies in, so we try all $2^n$ of them. 

We show that in fact sufficiently many samples from $P_{\vec x, r}$, where $\vec x \in 2\cL + \vec v^*$ is sufficiently short, suffice to find us our ``good weights.''
For \(0<\tau\le1/2\), consider the overlap
\[
I_\tau(P_{v^*,r},P_{x,r})
=
\int P_{\vec v^*,r}(\vec a)^{1-\tau}
P_{\vec x,r}(\vec a)^\tau\,d\vec a.
\]
Writing
\[
J_\tau(a,b)=
\bigl((1-\tau)a^2+\tau b^2\bigr)^{p/2}
-(1-\tau)|a|^p-\tau|b|^p,
\]
we show in \cref{lem:distribution_overlap} that 
\[
I_\tau(P_{u,r},P_{x,r})
=
\exp\left(
-\frac1{r^p}\sum_j J_\tau(u_j,x_j)
\right).
\]

A key observation in the $\cvp$ algorithm of \cite{ADS} is that the parallelogram identity relates the distance (from the target) of two lattice vectors in the same coset to the distance of their mid-point from the target.\footnote{
Indeed, for $F_2(\vec x)=\|\vec x\|_2^2$, the parallelogram identity gives
$$
\frac{F_2(\vec x)+F_2(\vec y)}{2}
=
F_2\left(\frac{\vec x+\vec y}{2}\right)
+F_2\paren{\frac{\vec x - \vec y}{2}}.
$$
In particular, if $\vec x,\vec y\in\cL-\vec t$ belong to the same coset modulo $2\cL$, their midpoint is also in $\cL-\vec t$. 
Thus, if $\vec y$ is optimal and $F_2(\vec x)\le F_2(\vec y)+\Delta$, then $F_2(\vec x-\vec y) \le 2\Delta$ and thus they are close.
This is the geometric observation underlying their clustering argument (see \cite[Lemma~5.1]{ADS}).
}
For $p\neq 2$, we no longer have such an identity.
Nevertheless, convexity still guarantees that 
\begin{equation}
\label{eqn:midpoint_convex}
    F_p\paren{\frac{\vec x + \vec y}{2}} \leq \frac{F_p(\vec x) + F_p(\vec y)}{2}.
\end{equation}

Using this fact, we are able to show that when $\norm{\vec x}_p^p$ is sufficiently close to $\norm{\vec v^*}_p^p$, $J$ is bounded as\footnote{This is exactly where we need the assumption that $p > 1$. Precisely, the algorithm works by being able to approximate the curvature of the $\ell_p$ metric using an $\ell_2$ metric. However, the most interesting case of $p = 1$ has no useful curvature we can exploit to locally approximate the $\ell_1$ objective.}  

\[
\sum_j J_\tau(v_j^*,x_j)
\le
O_p(\tau^{p/2})
\left[
\frac{F_p(\vec v^*)+F_p(\vec x)}2
-
F_p\left(\frac{\vec v^*+ \vec x}{2}\right)
\right] 
\le
O_p(\tau^{p/2})
\left[
F_p(\vec x)
-
F_p(\vec v^*)
\right]. 
\]
Therefore, 
\[ I_\tau(P_{\vec v^*,r},P_{\vec x,r}) \geq 2^{-\order{n}}, \]
and we show in \cref{lem:close_is_ok} that we can use the above techniques to find a progressively closer lattice point.

\subsection{Further related work}
\label{sec:related-work}

\paragraph{Integer programming and unbounded subset sum.}
Lenstra showed that integer programming in a fixed number of variables is
polynomial \cite{DBLP:journals/mor/Lenstra83}; Kannan improved the dependence
to $n^{O(n)}$ \cite{DBLP:conf/stoc/Kannan83,Kannan1987MinkowskisCB}, Frank and
Tardos showed how to reduce the size of the numbers involved
\cite{Frank1987AnAO}, Kannan and Lov\'asz studied the underlying flatness
phenomena through covering minima \cite{KL88}, Hildebrand and K\"oppe
extended the approach to quasiconvex polynomial objectives \cite{HK13},
Dadush obtained time $n^n2^{O(n)}$ for general convex bodies and a randomized
sieving framework for approximate integer programming
\cite{DadushThesis12,Dadush14}, Dadush, Eisenbrand and Rothvoss reduced exact
to approximate integer programming \cite{DER25}, and Reis and Rothvoss
reached $(\log2n)^{O(n)}$ \cite{DBLP:conf/focs/ReisR23}; see
\cite{Eisenbrand10} for a survey. Integer programs with few constraints and
small coefficients admit pseudopolynomial algorithms
\cite{Papadimitriou81,eisenbrand2019proximity,DBLP:conf/innovations/JansenR19}.
Whether integer programming can be solved in time $2^{O(n)}$ is open.

The endpoint $p=1$ connects exact $\cvp$ to this question through unbounded
subset sum: given positive integers $a_1,\dots,a_M$ and $T$, decide whether
$\sum_ia_iz_i=T$ has a solution $\vec z\in\Z_{\ge0}^M$. With $Q=T+1$, let
$B\in\Z^{(M+1)\times M}$ have rows $\diag(a_1,\dots,a_M)$ and
$(Qa_1,\dots,Qa_M)$, and let $\vec t=(0,\dots,0,QT)$. Then
$\norm{B\vec z-\vec t}_1=\sum_ia_i|z_i|+Q\,\bigl|\sum_ia_iz_i-T\bigr|$, so
$\dist_1(\vec t,\cL)\le T$ if and only if the instance is feasible, and an
algorithm for exact $\cvp_1$ running in time single-exponential in the
dimension would solve unbounded subset sum in time $2^{O(M)}$ times a
polynomial. That is open: the best general bound is the $(\log M)^{O(M)}$ of
\cite{DBLP:conf/focs/ReisR23}, polynomial-time algorithms are known when the
target is large \cite{aggarwal24,CMZ25}, special cases are treated in
\cite{DBLP:journals/corr/abs-2103-09080}, and pseudopolynomial algorithms
\cite{bocker2007fast,DBLP:conf/soda/Bringmann17,DBLP:conf/innovations/JansenR19,DBLP:conf/soda/Klein22,DBLP:journals/jcss/ChanH22}
are incomparable. The closely related Frobenius problem is NP-hard in general
and polynomial in fixed dimension
\cite{ramirez1996complexity,kannan1992lattice,Alfonsin2005TheDF}. Our $\cvp$
algorithm stops short of $p=1$, while our $\svp$ algorithm covers it.

A natural remaining open question is whether our techniques can be extended to work for $p=1$. 
Our algorithms crucially uses the fact the $\ell_p$ ball, for $1 < p < 2$, is strictly convex.
Due to this, an approximate $\cvp$ solution provides useful geometric information about its neighborhood that the algorithms exploit. 
This fails for $\ell_1$. 
However, the mixture identity holds for $\ell_1$ and has many nice properties one may exploit.


\section*{Acknowledgments}
This work was supported by NRF grant NRF-NRFI09-0005.
AM is supported by Singapore's NQSS scholarship.
AM thanks Li Zeyong, Saswata Mukherjee and Rishav Gupta for many helpful discussions. 

\section*{Statement of AI Use}
The initial idea to use the $\ell_2$ objective function to solve the $\ell_1$ problem via relating the corresponding (super)-Gaussian distributions was due to the authors. 
We found stable distributions from \cite{nolan2020} via a prior work on Bayesian learning, and eventually got an algorithm for $\svp_p$ for $1 \leq p < 2$ that was better than prior work. 
However, we were unable to make our ideas work for exact $\cvp$. Eventually the age of AI math came about, and after long deliberations on our previous attempts, GPT Sol and Astra helped us prove that a similar algorithm works for exact $\cvp$. We eventually learned that GPT pulled these techniques from the literature on Iteratively Weighted Least Squares and Dikin random walks, which are techniques invented to solve continuous optimization problems and sampling (see \cite{KV24,Gentle2024}). 
The version of the $\svp$ algorithm in the present draft is one we got after conversations with ChatGPT $5.6$ with a further improved constant in the exponent. 
All of the writing was by the authors. 

	\section{Preliminaries}
	\label{sec:prelims}
	\subsection{Notation and model of computation}
\label{sec:model}

For $\vec x \in \R^n$ and $p \ge 1$, let $\norm{\vec x}_p:=(\sum_j|x_j|^p)^{1/p}$ and we omit $p$, when $p=2$. For a function $f$ on $\R^n$ and a countable set $S$ we write $f(S):=\sum_{\vec x\in S}f(\vec x)$. We write $\ln$ for the natural logarithm and $\log$ for the logarithm to base $2$, and $|I|$ for the bit length of the input of the problem being solved.

The \emph{statistical distance} between two probability distributions $P$ and $Q$ on a set $\mathcal{S}$ is 
\[
\mathsf{SD}(P;Q) := \frac{1}{2} \sum_{x \in \mathcal{S}} |P(x) - Q(x)| \;.
\]
For any event $E$, we have $|P(E) - Q(E)| \le \mathsf{SD}(P;Q)$. 


\paragraph{Model of Computation.}

For simplicity of the exposition, we analyze our algorithms in the real-RAM model: a real number is stored exactly; arithmetic operations, comparisons, and evaluations of $\ln x$, $e^x$, $\sin x$ and $x^y$ take unit time; and a uniformly random real number in $[0,1]$ can be drawn in unit time. 
However, these results also hold in the word-RAM model with $\poly(\abs{I})$ overhead, where $\abs{I}$ is the size of the input, for every fixed, rational $p$ in $1 < p < 2$.
In particular, our algorithms use Euclidean $\svp_2$ and $\cvp_2$ subroutines on ``perturbed" lattices, which may have real entries. 
The following lemma shows that we can approximate the entries to $\poly(\abs{I})$ bits of precision, with a $\poly(\abs{I})$ overhead. 

\begin{lemma}
    Let $B \in \R^{n \times k}$ be the basis of a full column rank lattice with the minimum singular value $\sigma_{\mathrm{min}}(B)$.
    Suppose that for $b \in \N$, the matrix $\widehat B \in \Q^{n \times k}$ satisfies \[
    \max_{i,j}|B_{ij}-\widehat B_{ij}|\leq 2^{-b}.\]
    Then we have that for any $\vec z \in \Z^k$, \[ \norm{\widehat B \vec z - B \vec z}_2 \leq \frac{\sqrt{nk}2^{-b}}{\sigma_{\mathrm{min}}(B)} \norm{B\vec z}_2. \]
\end{lemma}

\begin{proof}
    By sub-multiplicativity of matrix-vector multiplication and the definition of singular value, \[ \norm{\widehat B \vec z - B \vec z}_2 \leq \sqrt{nk}2^{-b}\norm{\vec z}_2 \leq \frac{\sqrt{nk}2^{-b}}{\sigma_{\mathrm{min}}(B)} \norm{B\vec z}_2.\qedhere \]
\end{proof}

If our input lattice basis $B$ is rational, we have that \(
\sigma_{\min}(B)^{-1}\leq 2^{\poly(\abs{I})}.
\)
Therefore, by choosing $b$ appropriately, we can reach an exponentially small (in $\abs{I}$) distortion in the lengths of the lattice vectors. 
Moreover, the upper bounds in our proof of \cref{thm:cvp} are loose enough to recover exact $\cvp_p$ solutions from these rounded $\ell_2$ solutions. In particular, they are used to solve a variant of $\cvp_2$ in which the level sets are ellipsoids (see \cref{cor:ellipsoid_cvp}). 
Lastly, all the samplers from \cref{sec:3_dg_mix} can be implemented to return the samples up to desired precision (that is $\poly(\abs{I})$). 

The following inequality is used in \Cref{lem:distribution_overlap}; this is where $p>1$ is necessary.

\begin{lemma}
    \label{lem:j_upper_bound}
    Let $0 < \tau \leq 1/2$ and $1 < p < 2$. For $a, b \in \R$, define \[J_\tau(a,b) \coloneqq \bigl((1-\tau)a^2+\tau b^2\bigr)^{p/2} -(1-\tau)|a|^p-\tau|b|^p.\]
    Then, 
    \begin{equation}
        \label{eq:j_upper_bound}
        J_\tau(a,b) \le \frac{64}{p-1}\,\tau^{p/2} \left( \frac{|a|^p+|b|^p}{2} -\left|\frac{a+b}{2}\right|^p \right).
    \end{equation}
\end{lemma}

\begin{proof}
    The claim is immediate when $a=b=0$. Replacing $a,b$ by $|a|,|b|$ leaves $J_\tau$ unchanged and can only decrease the expression on the right side of \cref{eq:j_upper_bound}. Thus we may assume $a,b\ge0$ and write $M=\max(a,b)>0$.

    The function
    \[
        z \longmapsto z^p-\frac{p(p-1)}2M^{p-2}z^2
    \]
    is convex on $[0,M]$: its second derivative is $p(p-1)(z^{p-2}-M^{p-2})\ge0$ for $0<z\le M$, and its first derivative is continuous at zero. Applying convexity at $(a+b)/2$, we get
    \begin{equation}
        \label{eq:scalar_midpoint_lower}
        \frac{a^p+b^p}{2}-\left(\frac{a+b}{2}\right)^p
        \ge \frac{p(p-1)}8M^{p-2}(a-b)^2.
    \end{equation}
    We now consider two cases.

    \textbf{Case 1: $\min(a,b)\le M/2$.}
    Subadditivity of $z^{p/2}$ gives
    \[
        \begin{aligned}
            J_\tau(a,b)
            &\le \bigl((1-\tau)^{p/2}-(1-\tau)\bigr)a^p
                +\bigl(\tau^{p/2}-\tau\bigr)b^p\\
            &\le \tau^{p/2}(a^p+b^p)
            \le 2\tau^{p/2}M^p.
        \end{aligned}
    \]
    Since $|a-b|\ge M/2$, it follows that
    \begin{equation}
        \label{eq:scalar_far_upper}
        J_\tau(a,b)\le 8\tau^{p/2}M^{p-2}(a-b)^2.
    \end{equation}
    Combining \cref{eq:scalar_far_upper,eq:scalar_midpoint_lower} and using $1/p\le1$ proves the claim in this case.

    \textbf{Case 2: $\min(a,b)>M/2$.}
    If $a=b$, the claim is immediate, so assume $a\ne b$. Fix $a$ and, for $z>0$, define
    \[
        \begin{aligned}
            q(z)&\coloneqq (1-\tau)a^2+\tau z^2,\\
            f(z)&\coloneqq J_\tau(a,z)
                =q(z)^{p/2}-(1-\tau)a^p-\tau z^p.
        \end{aligned}
    \]
    Differentiating gives
    \[
        f'(z)=p\tau z\,q(z)^{p/2-1}-p\tau z^{p-1}.
    \]
    Since $q(a)=a^2$, we have
    \begin{equation}
        \label{eq:scalar_near_initial}
        f(a)=f'(a)=0.
    \end{equation}
    Differentiating once more,
    \[
        f''(z)=p\tau q(z)^{p/2-1}
        +p(p-2)\tau^2z^2q(z)^{p/2-2}
        -p(p-1)\tau z^{p-2}.
    \]
    The last two terms are nonpositive because $1<p<2$. For every $z$ between $a$ and $b$, both $a,z\ge M/2$, so
    \[
        q(z)\ge (1-\tau)\frac{M^2}{4}+\tau\frac{M^2}{4}
        =\frac{M^2}{4}.
    \]
    Since $p/2-1<0$, it follows that
    \begin{equation}
        \label{eq:scalar_second_derivative_upper}
        f''(z)\le p\tau q(z)^{p/2-1}
        \le p\tau\left(\frac{M^2}{4}\right)^{p/2-1}
        =p\,2^{2-p}\tau M^{p-2}.
    \end{equation}
    Taylor's theorem gives, for some $\xi$ between $a$ and $b$,
    \[
        f(b)=f(a)+f'(a)(b-a)+\frac12f''(\xi)(b-a)^2.
    \]
    Using \cref{eq:scalar_near_initial,eq:scalar_second_derivative_upper}, we obtain
    \begin{equation}
        \label{eq:scalar_near_upper}
        J_\tau(a,b)=f(b)
        \le p\,2^{1-p}\tau M^{p-2}(a-b)^2.
    \end{equation}
    Combining \cref{eq:scalar_near_upper,eq:scalar_midpoint_lower} and using $2^{1-p}\le1$ and $\tau\le\tau^{p/2}$ proves \cref{eq:j_upper_bound} in this case as well.
\end{proof}

\begin{theorem}[H\"older's Inequality; Theorem 3.5 from \cite{Rudin1987}]
    \label{thm:holder}
    Let $p,q > 1$ be such that \[ \frac{1}{p} + \frac{1}{q} = 1. \]
    Then, 
\[ \int_S |f(\vec s)g(\vec s)|\,d\vec s
\le
\left(\int_S |f(\vec s)|^p\,d\vec s\right)^{1/p}
\left(\int_S |g(\vec s)|^{q}\,d\vec s\right)^{1/q}. \]
\end{theorem}

\subsection{Stable Distributions}
\label{ssec:stabel_dist}

Notice that for $y \in [0, \infty)$ and $0 < p < 2$, the function \[ y \mapsto \exp\paren{-y^{p/2}} \] is completely monotone, that is, $(-1)^k\frac{d^k}{dy^k}\exp(-y^{p/2})\ge0$ for all $k\ge0$ and $y>0$.

Therefore, it follows from Bernstein's theorem \cite{Bernstein1929} (cf. \cite[Theorem 1.5]{Berg2007}) that this function is the Laplace transform of a positive measure $\mu_{p/2}$ on $[0, \infty)$, 
\[ \exp\paren{-y^{p/2}}
=
\int_0^\infty \exp\paren{-ay}\,d\mu_{p/2}(a). \]

It follows that \[ \mu_{p/2}([0, \infty)) = \lim_{y \to 0^+} \exp\paren{-y^{p/2}} = 1, \]
and therefore $\mu_{p/2}$ is a probability measure. 
It is well known that the corresponding random variable is the positive $p/2$-stable random variable \cite{nolan2020}.

Writing $\mu_{p/2}(da)=g_{p/2}(a)\,da$ for the density\footnote{It is known that the random variable admits a density, see \cite[Theorem 1.1]{nolan2020}.} of the $p/2$-stable random variable, we have that 

\begin{equation}
    \label{eqn:laplace_of_stabel_density}
    \exp\paren{- y^{p/2}} = \int_0^\infty \exp\paren{- a y} g_{p/2}(a) da.
\end{equation}

Substituting $y = (x/r)^2$ for $x \in \R$, $r > 0$ and $a = \pi r^2/s^2$, we get that  

\begin{equation}
    \label{eqn:one_variable_identity}
    \exp\paren{-\frac{\abs{x}^{p}}{r^p}}
=
\int_0^\infty \exp\paren{-\pi \frac{x^2}{s^2}} g_{p/2}\paren{\frac{\pi r^2}{s^2}} \frac{2\pi r^2}{s^3} ds.
\end{equation}

We now mention some facts about stable densities that will be useful throughout this work. 
Let \[
\alpha=\frac p2,\qquad
q=\frac{\alpha}{1-\alpha}=\frac p{2-p},\qquad
\kappa_p=(1-\alpha)\alpha^q,\qquad
\beta=2^{1/q}.
\]

Notice that we have 
\begin{equation}
    \label{eq:beta_is_2}
    \beta^q=2, \qquad (2\beta)^\alpha=2.
\end{equation}

For $0 < \theta < \pi$, define
    \[A_\alpha(\theta) \coloneqq
\frac{ 
\sin(\alpha\theta)^q\sin((1-\alpha)\theta)
}{
\sin(\theta)^{q+1}
}.\]

Zolotarev \cite{zolotarev} gave the following integral representation of the distribution function of the stable distribution (cf. \cite[Equation~(1)]{devroye}): for $a>0$,
\begin{equation}
    \label{eqn:zolotarev_stable}
    \int_0^a g_\alpha(u)\,du
=
\frac1\pi\int_0^\pi
e^{-A_\alpha(\theta)a^{-q}}\,d\theta.
\end{equation}

Differentiating \cref{eqn:zolotarev_stable} with respect to \(a\) we get that 
    \begin{equation}
        \label{eq:stab1}
        g_\alpha(a)
=
\frac{q}{\pi a^{q+1}}
\int_0^\pi
A_\alpha(\theta)e^{-A_\alpha(\theta)a^{-q}}\,d\theta.
    \end{equation}

    The representation \cref{eqn:zolotarev_stable} also gives a simple way to sample from the density $g_\alpha$, due to Kanter \cite{Kanter75} (see also \cite{CMS76}).
 
\begin{fact}
\label{fact:kanter}
Let $U$ be uniform on $(0,\pi)$, and let $E$ be a standard exponential random variable independent of $U$. Then $(A_\alpha(U)/E)^{1/q}$ has density $g_\alpha$.
\end{fact}
 
\begin{proof}
For $a>0$,
\[ \Pr\bigl[(A_\alpha(U)/E)^{1/q}\le a\bigr]=\Pr\bigl[E\ge A_\alpha(U)\,a^{-q}\bigr]=\expect_U\bigl[e^{-A_\alpha(U)a^{-q}}\bigr], \]
which is the right-hand side of \cref{eqn:zolotarev_stable}.
\end{proof}

We will need the following technical lemma. 

\begin{lemma}
    \label{lem:stabe_density_comparison}
    For $a > 0$, 
    \[ g_\alpha(a)\exp\left(\frac{\kappa_p a^{-q}}2\right) \le 2\beta\,g_\alpha(\beta a). \]
\end{lemma}

\begin{proof}
Since $\sin x$ for $x \in [0,\pi]$ is a concave function, we have that \[\sin(\alpha\theta)\ge\alpha\sin\theta,
\qquad
\sin((1-\alpha)\theta)\ge(1-\alpha)\sin\theta,\]
which implies that \[ A_\alpha(\theta) \geq \kappa_p \implies A_\alpha(\theta)-\kappa_p/2\ge A_\alpha(\theta)/2.\] 

Using \cref{eq:stab1} we get that 
    \[\begin{aligned}
    g_\alpha(a)e^{\kappa_p a^{-q}/2}
    &=
    \frac{q}{\pi a^{q+1}}
    \int_0^\pi
    A_\alpha(\theta)
    e^{-(A_\alpha(\theta)-\kappa_p/2)a^{-q}}\,d\theta\\
    &\le
    \frac{q}{\pi a^{q+1}}
    \int_0^\pi
    A_\alpha(\theta)e^{-A_\alpha(\theta)a^{-q}/2}\,d\theta \\ 
    &\leq 2\beta g_\alpha(\beta a),
    \end{aligned}
    \]
    where in the last inequality we have used \cref{eq:beta_is_2} and \cref{eq:stab1} a second time.  
\end{proof}

For $\lambda\ge0$, the \emph{exponentially tilted} stable distribution $G_{\alpha,\lambda}$ is the probability distribution on $(0,\infty)$ with density \[ \exp\paren{\lambda^\alpha - \lambda a} g_{\alpha}(a); \]
by \cref{eqn:laplace_of_stabel_density} this is a probability density, and $G_{\alpha,0}$ has density $g_\alpha$. Devroye \cite[Theorem 3]{devroye} gave an algorithm to sample from it.

\begin{lemma}[\cite{devroye}, Theorem 3]
    \label{lem:devroye_sample_stable}
   There is an algorithm that, given $\alpha \in (0, 1)$ and $\lambda \geq 0$, outputs a sample from $G_{\alpha,\lambda}$. Its uniform expected running time over all \(\alpha\in(0,1)\) and \(\lambda\ge0\) is $\mathcal{O}(1)$.
\end{lemma}

\subsection{Lattices}

A lattice of rank $k$ in $\R^n$ is a set $\cL=\cL(B):=B\Z^k$, where $B\in\R^{n\times k}$ has linearly independent columns $\vec b_1,\dots,\vec b_k$; the matrix $B$ is a \emph{basis} of $\cL$, and $\cL$ has \emph{full rank} if $k=n$. For $p\ge1$ and $\vec t\in\R^m$ we write
\[ \lambda_1^{(p)}(\cL):=\min_{\vec x\in\cL\setminus\{\vec 0\}}\norm{\vec x}_p
\qquad\text{and}\qquad
\dist_p(\cL,\vec t):=\min_{\vec x\in\cL}\norm{\vec x-\vec t}_p . \]
A vector $\vec x\in\cL$ with $\norm{\vec x-\vec t}_p=\dist_p(\cL,\vec t)$ is a \emph{closest vector} to $\vec t$.

\begin{fact}[Cosets of a Lattice]
\label{fact:parity}
Let $\cL=\cL(B)$ be a lattice of rank $n$.
\begin{enumerate}
    \item $\cL$ is the disjoint union of the $2^n$ cosets $2\cL+B\vec c$, $\vec c\in\{0,1\}^n$.
    \item If $\vec x$ and $\vec y$ lie in the same coset, then $\frac{\vec x+\vec y}{2}\in\cL$.
    \item If moreover $\cL\subseteq\Z^m$ and $\vec x\neq\vec y$, then $\norm{\vec x-\vec y}_2\ge2$.
\end{enumerate}
\end{fact}
 
\begin{proof}
(1) Write $\vec x=B\vec z$ with $\vec z\in\Z^n$ and reduce $\vec z$ modulo $2$; the cosets are disjoint because the columns of $B$ are linearly independent. (2) $\frac{\vec x+\vec y}{2}=\vec y+\frac{\vec x-\vec y}{2}$ and $\vec x-\vec y\in2\cL$. (3) $\vec x-\vec y$ is a nonzero vector of $2\cL\subseteq2\Z^m$, so one of its coordinates is a nonzero even integer.
\end{proof}

\textbf{Discrete Super-Gaussians.}
For $1\le p<2$, $r > 0$ and $\vec x \in \R^n$, define 
 \[ f_p(\vec x, r) \coloneqq \exp\paren{-\frac{\|\vec x\|_p^p}{r^p}}\;, \]
and for a lattice $\cL\subseteq\R^n$ and $\vec t\in\R^n$ let
 \[ f_p(\cL - \vec t, r) \coloneqq \sum_{\vec x \in \cL-\vec t} \exp\left(-\frac{\|\vec x\|_p^p}{r^p}\right) \;, \]
which is finite.
The \emph{$\ell_p$ discrete Gaussian} $D_p(\cL - \vec t, r)$ is the distribution on $\cL - \vec t$ given by
\begin{equation}
\label{eqn:d_p_def}
\Pr_{X \sim D_p(\cL-\vec t,r)}[ X = \vec x] = \frac{f_p(\vec x, r)}{f_p(\cL - \vec t, r)}\qquad\text{for }\vec x\in\cL-\vec t \;. 
\end{equation}

For $s>0$, the (Euclidean) discrete Gaussian $D_{\cL,s}$ is the distribution on $\cL$ that gives probability proportional to $e^{-\pi\norm{\vec x}^2/s^2}$ to $\vec x\in\cL$; it is $D_2(\cL,s/\sqrt\pi)$.
 
\subsection{Computational Problems and Algorithms.}
\label{sec:problems}

We are primarily interested in the following two optimization problems over lattices. 

\begin{definition}[Shortest Vector Problem, $\svp_p$]
    \label{def:svp}
    For $p\ge1$, the problem $\svp_p$ is: given a basis of a lattice $\cL \subseteq \Q^n$, output a vector $\vec x\in\cL\setminus\{\vec 0\}$ with $\norm{\vec x}_p=\lambda_1^{(p)}(\cL)$.
\end{definition}

\begin{definition}[Closest Vector Problem, $\cvp_p$]
    \label{def:cvp}
     For $p\ge1$, the problem $\cvp_p$ is: given a basis of a lattice $\cL \subseteq \Q^n$, a target $\vec t \in \Q^n$, output a vector $\vec x\in\cL$ with $\norm{\vec x - \vec t}_p=\min_{\vec y \in \cL} \norm{\vec y - \vec t}_p$.
\end{definition}

In the Euclidean norm both problems can be solved in time $2^{n+o(n)}\poly(|I|)$ \cite{ADRS,ADS}. We use the following two results.
 
\begin{theorem}[\cite{ADS}, Corollary 6.5]
    \label{thm:ads_cvp}
    There is a randomized algorithm that, given a basis of a lattice $\cL \subseteq \Q^n$ of rank $n$ and a target $\vec t\in\Q^n$, outputs a vector of $\cL$ which, with probability at least $1/2$, is a closest vector to $\vec t$ in the $\ell_2$ norm.\footnote{\cite[Corollary 6.5]{ADS} gives a constant success probability. Repeating the algorithm $k$ times and keeping the output closest to $\vec t$ raises it to $1-2^{-\Omega(k)}$; probability $1/2$ is all we use.} It runs in time $2^{n + o(n)}\poly(|I|)$.
\end{theorem}

Note that $\cvp$ can be equivalently defined as finding a {\em shortest vector in a lattice shift.} We will frequently think of this perspective.

We will use the following sampler from \cite{ADRS}.

\begin{theorem}[\cite{ADRS}, Theorem 3.7]
    \label{thm:adrs_sampler}
    There is an algorithm that, given a basis of a lattice $\cL \subseteq \Q^n$ of rank $n$, a parameter $s > 0$ and $\kappa \geq \Omega(n)$, outputs $2^{n/2}$ vectors of $\cL$ whose joint distribution is within statistical distance $e^{-\Omega(\kappa)}$ of the distribution of $2^{n/2}$ independent samples from $D_{\cL,s}$. It runs in time $2^{n + \polylog(\kappa) + o(n)}\poly(|I|)$.
\end{theorem}

In our $\cvp$ algorithm, the Euclidean $\cvp$ algorithm is applied to a coset of $2\cL$, in a diagonal metric.
 
\begin{corollary}
    \label{cor:ellipsoid_cvp}
    There is a randomized algorithm that, given a basis $B$ of a lattice $\cL$ of rank $n$, a vector $\vec w \in \R^n$ and widths $\vec s \in \R^n_{>0}$, outputs a vector of the coset $2\cL+\vec w$ which, with probability at least $1/2$, minimizes
    \[ \sum_{j=1}^n \frac{y_j^2}{s_{j}^2}\qquad\text{over }\vec y \in 2\cL+\vec w. \]
    It runs in time $2^{n + o(n)}\poly(|I|)$.
\end{corollary}



\begin{proof}
    Let $D:=\diag(1/s_1,\ldots,1/s_n)$. Every $\vec y\in2\cL+\vec w$ can be written as $\vec y=2B\vec z+\vec w$ with $\vec z\in\Z^n$, and then
    \[ \sum_{j=1}^n\frac{y_j^2}{s_j^2}=\norm{D\vec y}^2=\norm{2DB\vec z-(-D\vec w)}^2 . \]
    So $\vec y=2B\vec z+\vec w$ is a minimizer if and only if $2DB\vec z$ is a closest vector to $-D\vec w$ in the lattice $\cL(2DB)$, in the $\ell_2$ norm. Run the algorithm of \cref{thm:ads_cvp} on the basis $2DB$ and the target $-D\vec w$; if it returns $\vec v$, output $D^{-1}\vec v+\vec w$. (If $\vec v\notin\cL(2DB)$, which can be checked by solving a linear system, output $\vec w$ instead.)
\end{proof}
 
With the deterministic algorithm of \cite{MV13} in place of \cref{thm:ads_cvp}, the algorithm of \cref{cor:ellipsoid_cvp} always outputs a minimizer, and runs in time $2^{2n+o(n)}\poly(|I|)$.

    \section{Discrete Gaussian Mixtures}
    \label{sec:3_dg_mix}
    In this section $p\in[1,2)$ and $\alpha=p/2$. We show that $D_p(\cL - \vec t, r)$, defined in \cref{eqn:d_p_def}, is a mixture of Euclidean discrete Gaussians. Following \cref{eqn:one_variable_identity}, consider a pair of random variables $(X,S)$ with values in $(\cL - \vec t) \times \R^n_{> 0}$ and joint density

\begin{equation}
    \label{eqn:joint_dist}
    \Pr[X=\vec x,\ S\in d\vec s] =
\frac{\exp\left(-\pi\sum_{j=1}^n x_j^2/s_j^2\right)}
     {f_p(\mathcal L - \vec t,r)}
\prod_{j=1}^n
\left[
g_{p/2}\!\left(\frac{\pi r^2}{s_j^2}\right)
\frac{2\pi r^2}{s_j^3}\,ds_j
\right].
\end{equation}

By \cref{eqn:one_variable_identity}, applied to each coordinate, the marginal distribution of $X$ is exactly $D_p(\cL - \vec t, r)$.

For $\vec s \in \R^n_{> 0}$ and $r > 0$, define $$h_p(\vec s, r) \coloneqq \prod_{j=1}^n
g_{p/2}\!\left(\frac{\pi r^2}{s_j^2}\right)
\frac{2\pi r^2}{s_j^3},$$ 

and for $\vec x \in \R^n$, define 

\[ \rho(\vec x, \vec s) \coloneqq \exp\!\left(-\pi\sum_j\frac{x_j^2}{s_j^2}\right). \]

Given $S=\vec s$, the distribution of $X$ is the anisotropic Euclidean discrete Gaussian
\begin{equation}
    \label{eqn:x_given_s}
    \Pr[X=\vec x\mid S=\vec s] =
    \frac{\rho(\vec x, \vec s)}
    {\rho(\cL - \vec t, \vec s)}.
\end{equation}

Using \cref{eqn:one_variable_identity}, we get the following identity
\begin{equation}
    \label{eqn:fp_as_marginal}
    f_p(\vec x, r) = \int_{\R^n_{>0}} \rho(\vec x, \vec s) h_p(\vec s, r) d\vec s \;.
\end{equation}

Therefore, the conditional density of $S$ given $X = \vec x$ is
\begin{equation}
    \label{eq:s_given_x}
    P_{\vec x, r}(\vec s) \coloneqq \frac{\rho(\vec x,\vec s)\,h_p(\vec s,r)}{f_p(\vec x,r)},
\end{equation}
and the marginal density of $S$ is
\begin{equation}
    \label{eqn:s_marginal}
    \frac{h_p(\vec s, r)\,\rho(\mathcal L-\vec t, \vec s)}{f_p(\mathcal L-\vec t,r)} .
\end{equation}

We call $h_p(\cdot,r)$ the \emph{prior} and $P_{\vec x,r}$ the \emph{posterior} at $\vec x$. The posterior is defined by \cref{eq:s_given_x} for every $\vec x\in\R^n$, and depends only on $\vec x$ and $r$, not on $\cL$ or $\vec t$. In terms of the precisions $\pi r^2/s_j^2$ it is a product of tilted stable distributions. Indeed, \cref{eq:s_given_x} is a product of one factor for each coordinate, and with $a_j:=\pi r^2/s_j^2$ and $\lambda_j:=x_j^2/r^2$ we have $\pi x_j^2/s_j^2=\lambda_ja_j$ and $\abs{x_j}^p/r^p=\lambda_j^\alpha$; so the change of variables $a_j=\pi r^2/s_j^2$ turns the $j$-th factor into the density $e^{\lambda_j^\alpha-\lambda_ja_j}g_\alpha(a_j)$ of $G_{\alpha,\lambda_j}$. Therefore,
\begin{equation}
    \label{eq:posterior_precision}
    \text{if }\vec s\sim P_{\vec x,r},\text{ then }\frac{\pi r^2}{s_1^2},\dots,\frac{\pi r^2}{s_n^2}\text{ are independent, and }\frac{\pi r^2}{s_j^2}\sim G_{\alpha,\,x_j^2/r^2}.
\end{equation}

\begin{lemma}[Sampling the posterior]
    \label{lem:sample_posterior}
    There is an algorithm that, given $\vec x \in \R^n$ and $r > 0$, outputs a sample from the density $P_{\vec x, r}$, in expected time $O(n)$.
\end{lemma}
 
\begin{proof}
    For $j=1,\dots,n$ independently, sample $a_j$ from $G_{\alpha,x_j^2/r^2}$ with \Cref{lem:devroye_sample_stable}, and output $\vec s$ with $s_j:=r\sqrt{\pi/a_j}$. By \cref{eq:posterior_precision}, $\vec s$ has density $P_{\vec x,r}$.
\end{proof}

\begin{lemma}[Sampling the prior]
    \label{lem:sample_hpr}
    There is an algorithm that, given $n$ and $r > 0$, outputs a sample from the density $h_p(\cdot, r)$ on $\R^n_{>0}$, in time $O(n)$.
\end{lemma}
 
\begin{proof}
    For $j \in [n]$, sample $a_j$ independently from $G_\alpha$ with \Cref{fact:kanter}, and output $\vec s$ with $s_j = r\sqrt{\pi/a_j}$. This is correct by \cref{eq:posterior_precision} with $\vec x=\vec 0$.
\end{proof}

\begin{remark}
    \label{rem:no_dgs}
    We never sample from $D_p(\cL - \vec t, r)$, nor from the marginal \cref{eqn:s_marginal} of $S$, which contains the lattice sum $\rho(\cL-\vec t,\vec s)$. For small $r$, the distribution $D_p(\cL - \vec t, r)$ is concentrated on the shortest vectors of $\cL-\vec t$, so a sampler for it would solve $\cvp_p$. Our algorithms sample only from the prior (\Cref{sec:svp}) and from posteriors (\Cref{sec:4_cvp}), which do not depend on the lattice, and use the lattice only through Euclidean algorithms in the diagonal metric that was sampled.
\end{remark}

    \section{\texorpdfstring{$2^{O(n)}$ Algorithm for $\cvp_p$}{Algorithm for CVP in ellp}}
    \label{sec:4_cvp}
    In this section $p\in(1,2)$ is fixed. By multiplying the basis and the target by a common denominator of the entries of the basis, which changes the input length only polynomially and multiplies all distances by the same factor, we may assume that $\cL\subseteq\Z^n$; this is used only in \Cref{lem:close_enough_is_best}. We prove the following theorem.

\begin{algorithm}[H]
    \caption{$\cvp_p$ for $p \in (1, 2)$}
    \label{alg:cvp}
    \algorithmicrequire{Lattice basis $B \in \Z^{n \times n}$; Target $\vec t \in \R^n$; Parameters $T$, $N$.}
    
    \algorithmicensure{$\vec v^* = \cvp_p(\cL(B), \vec t)$.}

    \begin{algorithmic}[1]
        \State $r_1 = n \cdot \max_j \norm{\vec b_j}_p + \norm{\vec t}_p$
        \For{$\vec c \in \set{0, 1}^n$}\label{aln:loop_c}
            \State $\vec x_{\vec c}^{(1)} = B \vec c$
            \For{$i \in [T]$}\label{aln:loop_T}
            \State $r_i = r_1/2^{i-1}$
                \For{$j \in [N]$}\label{aln:loop_N}
                    \State Sample $\displaystyle{\vec s_j \sim P_{\vec x_{\vec c}^{(i)} - \vec t, r_i}}$\label{aln:sample_s}
                    \State Solve $\displaystyle{\vec y_{j} \gets \argmin_{\vec y \in 2\cL + \vec x_{\vec c}^{(i)} - \vec t} \paren{\sum_{k=1}^n \frac{y_k^2}{s_{j,k}^2}}}$\Comment{Use the algorithm from \cref{cor:ellipsoid_cvp}}\label{aln:ell2_opt}
                \EndFor
               \State Let $\vec w$ be a vector of least $\ell_p$ norm in $\set{\vec x_{\vec c}^{(i)} - \vec t, \vec y_1, \ldots \vec y_N} \subset \cL - \vec t$
                \State Set $\vec x_{\vec c}^{(i+1)} = \vec w + \vec t \in 2\cL + B\vec c$
            \EndFor
        \EndFor
        \State \Return the vector $\vec x_{\vec c}^{(T+1)}$ with the least value of $\norm{\vec x_{\vec c}^{(T+1)}-\vec t}_p$ over $\vec c\in\{0,1\}^n$
 
    \end{algorithmic}
\end{algorithm}

\begin{theorem}
    \label{thm:cvp}
    Let $p \in (1, 2)$. Given a basis of a lattice $\cL\subseteq\Q^n$ of rank $n$ and a target $\vec t \in \Q^n$, \Cref{alg:cvp} (after the scaling above) outputs a vector of $\cL$ closest to $\vec t$ in the $\ell_p$ norm with probability at least $e^{-1}$, in expected time $2^{(2+C_p)n+o(n)}\poly(|I|)$, where $C_p$ is defined below.
\end{theorem}
 
Running the algorithm $\ceil{e\ln(1/\delta)}$ times independently and returning the output closest to $\vec t$ gives a closest vector with probability at least $1-\delta$, which proves \Cref{thm:main-cvp}. The rest of this section will prove \Cref{thm:cvp}.

Let $\displaystyle{C_p \coloneqq 2 \paren{\frac{2048}{p-1}}^{2/p}}$ and $\vec v^* \in \cvp_p(\cL(B), \vec t)$.
Set $$T = 1 + \ceil{\frac{1}{p} \log_2\paren{\frac{16 n r_1^2}{p(p-1)}}}$$
and 
$$N = \ceil{2^{C_p n + 1} \log(T + 1)}.$$

Consider the coset $M \coloneqq 2\cL + \vec v^* - \vec t$, and let $d \coloneqq \dist_p(\cL, \vec t)$.
We will show that the algorithm finds a closest vector in the iteration of line \ref{aln:loop_c} in which $B\vec c \in 2\cL+\vec v^*$; in that iteration, every $\vec x_{\vec c}^{(i)}$ lies in $2\cL+\vec v^*$, and the set $2\cL + \vec x_{\vec c}^{(i)} - \vec t$ in line \ref{aln:ell2_opt} is $M$.
 
The algorithm works by iteratively reducing the distance of $\vec x_{\vec c}$ from the target. The proof consists of three steps.
For $\vec s\in\R^n_{>0}$, let $Y(\vec s)\subseteq M$ be the set of minimizers of $\sum_k y_k^2/s_k^2$ over $M$; when the call to \Cref{cor:ellipsoid_cvp} in line \ref{aln:ell2_opt} succeeds, it returns an element of $Y(\vec s_j)$.
First, we show that if $\vec x_{\vec c}$ were the shortest vector $\vec v^* - \vec t$, then with high probability every $\vec y\in Y(S)$ has $\norm{\vec y}_p^p$ only slightly larger than $d^p$.
 
Next, we show that the densities of $P_{\vec v^* - \vec t, r}$ and $P_{\vec x - \vec t, r}$ have a large overlap when $\vec x$ is {\em close enough} to $\vec t$. 
Finally we combine these results to conclude that the algorithm successfully reduces the distance of the current vector $\vec x_{\vec c}$ with a large enough probability. 
We will now prove these three results, followed by the proof of \cref{thm:cvp}.

\begin{lemma}
    \label{lem:closest_is_best}
    For any $r > 0$, we have that 
    \[ \Pr_{S \sim P_{\vec v^* - \vec t, r}}\bigl[\norm{\vec y}_p^p - d^p \leq 8nr^p\text{ for every }\vec y\in Y(S)\bigr] \geq 1 - 2^n \cdot e^{-4n}\;. \]
\end{lemma}

\begin{proof}
    Fix a coordinate $j \in [n]$, and write $u = v_j^* - t_j$. 
    For a fixed $s > 0$, consider the function \[ f(z) \coloneqq \abs{z}^p - \frac{\pi r^p}{s^2} z^2,\]
    which represents the gap between the quadratic objective solved in line \ref{aln:ell2_opt} and the desired objective. 
    One can check that $f$ is maximized at \[ z_{\max} \coloneqq \paren{\frac{ps^2}{2\pi r^p}}^{1/(2-p)} \]
    with \[ f(z_{\max}) = \left(1-\frac p2\right)
    \left(\frac{ps^2}{2\pi r^p}\right)^{p/(2-p)}. \]
    Consequently, for every $z \in \R$, 
    \begin{equation}
        \label{eq:closest_is_best_1}
        |z|^p
        \le
        \frac{\pi r^p}{s^2}z^2
        +
        \left(1-\frac p2\right)
        \left(\frac{ps^2}{2\pi r^p}\right)^{p/(2-p)}.
    \end{equation}

    Define \[ \Delta_{j,r}(s) \coloneqq \frac{\pi r^p u^2}{s^2}
    +
    \left(1-\frac p2\right)
    \left(\frac{ps^2}{2\pi r^p}\right)^{p/(2-p)}
    -|u|^p.\]

    Substituting this in \cref{eq:closest_is_best_1} and rearranging, we have that 
    \[|z|^p-|u|^p \le \frac{\pi r^p}{s^2}(z^2-u^2)+\Delta_{j,r}(s).\]

        Let $\vec y\in Y(S)$. Substituting $z = y_j$ and summing over $j$, we get that
    \begin{align*}
        \|\vec y\|_p^p-d^p &\le \pi r^p\sum_j \frac{y_j^2-(v_j^*-t_j)^2}{S_j^2} +\sum_j\Delta_{j,r}(S_j)\\
            &\le \sum_j\Delta_{j,r}(S_j),
    \end{align*}
    where the last inequality holds because $\vec v^*-\vec t\in M$ and $\vec y$ minimizes $\sum_k y_k^2/S_k^2$ over $M$.

    Next, we have the following claim. 
    \begin{claim}
        \label{cl:tangent_at_optimal}
        \[ \expect_{S \sim P_{\vec v^* - \vec t, r}}\left[\exp\paren{\frac{\Delta_{j,r}(S_j)}{2r^p}}\right] \leq 2. \]
    \end{claim}

\begin{proof}[Proof of \cref{cl:tangent_at_optimal}]
    Fix a coordinate $j \in [n]$, and let $u = v_j^* - t_j$ and $a = \pi r^2 / s^2$. 
    From \cref{eq:s_given_x} we have that in one dimension, 
    \[P_{u,r}(s)
        =
    \frac{e^{-\pi u^2/s^2}h_p(s,r)}
     {e^{-|u|^p/r^p}} = \frac{e^{-\pi u^2/s^2}}
     {e^{-|u|^p/r^p}} \cdot g_\alpha\left(\frac{\pi r^2}{s^2}\right)
    \frac{2\pi r^2}{s^3}.\]

    The conditional density of $a$ is \[ g_\alpha(a)
    \exp\left(-\frac{u^2}{r^2}a+\frac{|u|^p}{r^p}\right). \]

    Moreover, \[\frac{\Delta_{j,r}(s)}{r^p} = \frac{u^2}{r^2}a+\kappa_p a^{-q}-\frac{|u|^p}{r^p},\]
    where in the middle term we have used the fact that $p(q+1) = 2q$. 

    Therefore, 
    \[\begin{aligned}
    \expect \left[e^{\Delta_{j,r}(S_j)/(2r^p)}\right]
    &=
    e^{|u|^p/(2r^p)}
    \int_0^\infty
    g_\alpha(a)
    e^{-u^2a/(2r^2)+\kappa_p a^{-q}/2}\,da\\
    &\le
    2\beta\,e^{|u|^p/(2r^p)}
    \int_0^\infty
    g_\alpha(\beta a)e^{-u^2a/(2r^2)}\,da, && (\text{\cref{lem:stabe_density_comparison}}) \\ 
      &= 2 \exp\paren{\frac{\abs{u}^p}{2r^p}} \exp\paren{-\frac{\abs{u}^p}{(2\beta)^{p/2}r^p}} && (\text{\cref{eqn:laplace_of_stabel_density}}) \\ 
    &= 2 && \text{(since } (2\beta)^{p/2} = 2).
    \end{aligned}
    \]
    This ends the proof of the claim. 
\end{proof}

    Hence, if some $\vec y\in Y(S)$ has $\norm{\vec y}_p^p-d^p>8nr^p$, then $\sum_j\Delta_{j,r}(S_j)>8nr^p$. Now, by \cref{eq:posterior_precision} the coordinates of $S\sim P_{\vec v^*-\vec t,r}$ are independent, so by Markov's inequality and \cref{cl:tangent_at_optimal},
    \begin{align*}
        \Pr_{S \sim P_{\vec v^* - \vec t, r}}\left[\sum_j\Delta_{j,r}(S_j) > 8nr^p\right] &= \Pr\left[\prod_j e^{\Delta_{j,r}(S_j)/(2r^p)} > e^{4n}\right] \leq e^{-4n}\prod_j\expect\left[e^{\Delta_{j,r}(S_j)/(2r^p)}\right] \leq 2^n e^{-4n}.
    \end{align*}
\end{proof}

\begin{lemma}
    \label{lem:distribution_overlap}
       Let $0 < \tau \leq 1/2$ and let $\vec x \in 2\cL + \vec v^*$ be such that $\norm{\vec x - \vec t}_p^p \leq d^p + 8n(2r)^p$.
    We have that \[ I_\tau \coloneqq \int P_{\vec v^* - \vec t,r}(\vec s)^{1-\tau} P_{\vec x - \vec t, r}(\vec s)^\tau d\vec s \geq \exp\paren{ - \frac{1024 \tau^{p/2}n}{p-1} }. \]
 
\end{lemma}

\begin{proof}
    Using \cref{eq:s_given_x}, we have that \[ \begin{aligned}
        I_{\tau} &= \exp\paren{\frac{(1-\tau)d^p + \tau \norm{\vec x - \vec t}_p^p}{r^p}} \\ 
        &\quad\quad \times \int \exp\paren{ -\pi \sum_j \paren{ \frac{(1-\tau) (v_j^* - t_j)^2 + \tau (x_j - t_j)^2}{s_j^2} } } h_p(\vec s, r) d\vec s \\ 
        &= \exp\paren{- \frac{\sum_j J_{\tau}(v^*_j - t_j, x_j - t_j)}{r^p}},
    \end{aligned} \]
where we have used \cref{eqn:fp_as_marginal} in the last equality, and $J$ is as defined in \cref{lem:j_upper_bound}. 

    Recall that since $\vec x - \vec t, \vec v^* - \vec t \in M$, we have that \[ \frac{\vec x + \vec v^*}{2} \in \cL, \]
    and therefore 
    \begin{equation}
        \label{eqn:dist_comparison_same_coset}
        \norm{\frac{\vec x + \vec v^*}{2} - \vec t}_p^p \geq d^p.
    \end{equation}
    Using \cref{lem:j_upper_bound}, 
    \begin{align}
        \sum_j J_{\tau}(v_j^* - t_j, x_j - t_j) &\leq \frac{64 \tau^{p/2}}{p-1} \sum_j \paren{ \frac{ \abs{x_j - t_j}^p + \abs{v_j^* - t_j}^p }{2} - \abs{ \frac{ x_j + v_j^* }{2} - t_j }^p } \notag \\ 
        &= \frac{64 \tau^{p/2}}{p-1} \paren{ \frac{ \norm{\vec x - \vec t}^p_p + \norm{\vec v^* - \vec t}^p_p }{2} - \norm{ \frac{ \vec x + \vec v^* }{2} - \vec t }^p_p } \notag \\ 
        &\leq \frac{32 \tau^{p/2}}{p-1} \paren{ \norm{\vec x - \vec t}^p_p - d^p } \label{eqn:interm1}\\ 
        &\leq \frac{1024 \tau^{p/2}nr^p}{p-1} \label{eqn:interm2},
    \end{align}
    where in \cref{eqn:interm1} we have used \cref{eqn:dist_comparison_same_coset} and in \cref{eqn:interm2} we have used the assumption $\norm{\vec x - \vec t}_p^p \leq d^p + 8n(2r)^p$ and the fact that for $p \in (1,2)$, $2^p \leq 4$. The claim follows by substituting this in the definition of $I_\tau$. 
\end{proof}

\begin{lemma}
    \label{lem:close_is_ok}
     For any $r > 0$ and $\vec x \in 2\cL+\vec v^*$ such that $\norm{\vec x-\vec t}_p^p \leq d^p + 8n(2r)^p$, we have that
    \[ \Pr_{S \sim P_{\vec x - \vec t, r}}[\norm{\vec y}_p^p \leq d^p + 8nr^p\text{ for every }\vec y\in Y(S)] \geq 2^{-C_p n}. \]
 
\end{lemma}

\begin{proof}
    Set $\tau_p = 2/C_p$ in \cref{lem:distribution_overlap} and notice that 
    \begin{equation}
        \label{eq:close_is_ok1}
        I_{\tau_p} \geq e^{-n/2}.
    \end{equation}
     Let \[ E \coloneqq \set{ \vec s : \norm{\vec y}_p^p \leq d^p + 8nr^p\text{ for every }\vec y\in Y(\vec s) }. \]
    From \cref{lem:closest_is_best} we know that
    \begin{equation}
        \label{eq:close_is_ok2}
        P_{\vec v^* - \vec t, r}(E^c) \leq \exp\paren{- \paren{4 - \ln 2}n}.
    \end{equation}

    Next, recall the definition of $I_\tau$ from \cref{lem:distribution_overlap}. 
    Split the integral over $E$ and $E^c$, and use H\"older's inequality (\cref{thm:holder}) with $f = P_{\vec v^* - \vec t, r}$, $g = P_{\vec x - \vec t, r}$ $p = 1/(1-\tau)$, $q = 1/\tau$ to get 
    \begin{align}
        I_\tau &\leq P_{\vec v^* - \vec t, r}(E)^{1 - \tau} \cdot P_{\vec x - \vec t, r}(E)^{\tau} + P_{\vec v^* - \vec t, r}(E^c)^{1 - \tau} \cdot P_{\vec x - \vec t, r}(E^c)^{\tau} \notag \\ 
        &\leq P_{\vec x - \vec t, r}(E)^{\tau} + P_{\vec v^* - \vec t, r}(E^c)^{1 - \tau}\label{eq:close_is_ok3}.
    \end{align}

    Combining \cref{eq:close_is_ok1,eq:close_is_ok2,eq:close_is_ok3} and setting $\tau = \tau_p\le\frac12$, we get $P_{\vec x - \vec t, r}(E)^{\tau_p}\ge e^{-n/2}-e^{-(4-\ln2)n/2}\ge\frac12e^{-n/2}$, and hence
    \[ P_{\vec x - \vec t, r}(E) \geq \paren{\tfrac12e^{-n/2}}^{C_p/2}=2^{-C_p/2}e^{-C_pn/4}\ge2^{-C_p n}, \]
    where the last inequality holds for all $n\ge1$ because $n\ln2-\frac n4\ge\frac{\ln2}2$.
\end{proof}

\begin{lemma}
    \label{lem:close_enough_is_best}
    Let $p \in (1,2)$, $\cL \subset \Z^n$ be a lattice and $\vec t \in \R^n$. 
    Consider any $\vec x \in 2\cL + \vec v^*$ such that $\dist_p(\cL, \vec t) < \norm{\vec x - \vec t}_p$. 
    Let $r \in \R$ be such that \( \norm{\vec x - \vec t}_p \leq r\) and \(\dist_p(\cL, \vec t) \leq r. \)
    Then, \[ \norm{\vec x - \vec t}_p^p \geq \dist_p(\cL, \vec t)^p + p (p-1) r^{p-2}. \]
\end{lemma}

\begin{proof}
    Let $\vec v^* \in \cL$ be such that $\norm{\vec v^* - \vec t}_p = \dist_p(\cL, \vec t)$.
    For $-r \leq z \leq r$, consider the function \[ z \longmapsto  \abs{z}^p - \frac{p(p-1)}{2} r^{p-2}z^2.\]
    This function is convex, and therefore we have that for $z_1, z_2 \in [-r, r]$,  
    \[\frac{\abs{z_1}^p + \abs{z_2}^p}{2} - \abs{\frac{z_1 + z_2}{2}}^p \geq \frac{p(p-1)r^{p-2}}{8} \paren{z_1 - z_2}^2. \]

    Using this coordinatewise, we have that
    \[\begin{aligned}
    &\frac{\|\vec x-\vec t\|_p^p +\dist_p(\cL, \vec t)^p}{2}
    -\left\|\frac{\vec x+\vec v^*}{2}-\vec t\right\|_p^p\\
    &\hspace{2cm}\ge
    \frac{p(p-1)}8r^{p-2}\|\vec x-\vec v^*\|_2^2.
    \end{aligned}
    \]
    Since $\vec x - \vec v^* \in 2\cL$, we have that $\norm{\vec x - \vec v^*}_2^2 \geq 4$ and therefore 
    \begin{align*}
        \|\vec x-\vec t\|_p^p-d^p
    &\ge
    \frac{p(p-1)}4r^{p-2}
    \|\vec x-\vec v^*\|_2^2 \\ 
    &\geq  p(p-1)r^{p-2},
    \end{align*}
    which implies the claim.
\end{proof}

We are now ready to prove the main result. 

\begin{proof}[Proof of \cref{thm:cvp}]

  \textbf{Correctness.}
    Let $\vec c$ be such that $B\vec c \in 2\cL + \vec v^*$. We will show that the algorithm succeeds in this iteration of line \ref{aln:loop_c}.
    Suppose that at the start of the $i$th iteration of the loop in line \ref{aln:loop_T} we have $\norm{\vec x_{\vec c}^{(i)}-\vec t}_p^p\le d^p+8n(2r_i)^p$; for $i=1$ this holds because $\norm{B\vec c-\vec t}_p\le r_1$. Then for any $j$ in line \ref{aln:loop_N}, by \cref{lem:close_is_ok} with $r=r_i$ and $\vec x=\vec x_{\vec c}^{(i)}$, and since the call to \cref{cor:ellipsoid_cvp} succeeds with probability at least $1/2$ independently of $\vec s_j$,
    \[ \Pr[\norm{\vec y_j}_p^p \leq d^p + 8nr_{i}^p] \geq \tfrac12\cdot2^{-C_p n}. \]
    The $N$ samples are independent (conditioned on the priors), so
    \begin{align*}
        \Pr[\forall j \in [N], \norm{\vec y_j}_p^p > d^p + 8nr_{i}^p] &\leq (1-2^{-C_p n-1})^{N} \\
        &\leq \exp\paren{-N \cdot 2^{-C_p n-1}} \\
        &\leq \frac{1}{T+1},
    \end{align*}
    and if some $\vec y_j$ satisfies $\norm{\vec y_j}_p^p \leq d^p + 8nr_i^p = d^p + 8n(2r_{i+1})^p$, then so does $\vec x_{\vec c}^{(i+1)}-\vec t$, which is the shortest among the candidates. By induction over the $T$ iterations,
    \begin{align}
        \label{eq:final_1}
        \Pr\left[\norm{\vec x_{\vec c}^{(T+1)}-\vec t}_p^p \leq d^p + 8nr_{T}^p\right] &\geq \paren{1 - \frac{1}{T+1}}^{T} \geq e^{-1}.
    \end{align}

    After $T$ iterations, we have that 
    \begin{align}
        8nr^p_T &= 8nr^p_1 \cdot 2^{-p(T-1)} \notag \\ 
        &\leq \frac{1}{2} p(p-1) r_1^{p-2}\label{eqn:final_2}
    \end{align}
    by our choice of $T$.

      Since the candidate retained in each iteration is the shortest, $\norm{\vec x_{\vec c}^{(T+1)}-\vec t}_p\le\norm{B\vec c-\vec t}_p\le r_1$, and $d\le\norm{\vec t}_p\le r_1$. So \cref{eqn:final_2} and \cref{lem:close_enough_is_best} with $r=r_1$, together with \cref{eq:final_1}, imply that with probability at least $e^{-1}$, $\vec x_{\vec c}^{(T+1)}$ is a closest vector to $\vec t$; and then the vector returned, which is at least as close to $\vec t$, is a closest vector as well.

  \textbf{Complexity.}
    After the scaling, $r_1\ge1$ and $\log r_1\le\poly(|I|)$, so $T=\mathcal O_p(\log n+\log r_1)\le\poly(|I|)$.
    Each iteration of line \ref{aln:loop_N} samples $\vec s_j$ in expected time $O(n)$ (\cref{lem:sample_posterior}) and calls the algorithm of \cref{cor:ellipsoid_cvp}, which takes time $2^{n+o(n)}\poly(|I|)$.
    Therefore the expected running time of the algorithm is
    \[ 2^n \cdot T \cdot N \cdot 2^{n + o(n)}\poly(|I|) = 2^{(2+C_p)n+o(n)}\poly(|I|) \]
    as claimed.
 
\end{proof}

\begin{remark}[Lattices that do not have full rank]
    \label{rem:rank}
    Everything in this section goes through for a lattice $\cL\subseteq\Z^n$ of rank $k<n$, with a basis $B\in\Z^{n\times k}$, and any target $\vec t\in\Q^n$, not necessarily in the span of $\cL$. There are then $2^k$ cosets of $2\cL$ in $\cL$, and \Cref{cor:ellipsoid_cvp} is applied to lattices of rank $k$. But the widths $\vec s$ still have one coordinate for each of the $n$ coordinates of $\R^n$, and the bounds of \Cref{lem:closest_is_best,lem:distribution_overlap,lem:close_is_ok} are unchanged, because they are coordinate-wise products. So the running time is $2^{2k+C_pn+o(n)}\poly(|I|)\log(2/\delta)$, which is single-exponential in the dimension $n$.
\end{remark}


\bibliographystyle{alpha}
\bibliography{References.bib}

\newpage
\appendix

 \section{\texorpdfstring{$2^{O(k)}$ Algorithm for $\svp_p$}{Algorithm for SVP in ellp}}
\label{sec:svp}

We will need the following bound on the super-Gaussian mass of the lattice. 

\begin{lemma}
    \label{lem:upper_bound_fpl}
 Let $p \in [1, 2)$, let $\cL\subseteq\R^n$ be a lattice of rank $k$, and let $r > 0$. Then
    \[
f_p(\mathcal L,r)
\le
1+
2^k k!
\left(\frac{r}{\lambda_1^{(p)}}\right)^{pk}
\exp\left(\frac{(\lambda_1^{(p)})^p}{2r^p}\right)\;, \]
where $\lambda_1^{(p)}=\lambda_1^{(p)}(\cL)$.
\end{lemma}

\begin{proof}
    For $u \geq 1$, let \[ N(u) \coloneqq \abs{\set{ \vec y\in\mathcal L\setminus\{\vec 0\}: \|\vec y\|_p\le u\lambdaonep }}. \]
 
   The open $\ell_p$ balls of radius $\lambdaonep/2$ around the points of $\cL$ are pairwise disjoint, and those around $\vec 0$ and around the $N(u)$ points counted are contained in the ball $\cB_p^n(\vec 0, (u + 1 / 2)\lambda_1^{(p)})$. Comparing volumes in $\operatorname{span}(\cL)$,
    \begin{align}
        \label{eqn:packing}
        N(u) &\leq \frac{\vol_k\paren{\cB_p^n(\vec 0, (u + 1 / 2)\lambda_1^{(p)}) \cap \operatorname{span}(\cL)}}{\vol_k\paren{\cB_p^n(\vec 0, \lambda_1^{(p)} / 2) \cap \operatorname{span}(\cL)}} \leq (1 + 2u)^k.
    \end{align}

    Consider the function \[ g(u) \coloneqq \exp\paren{ - \paren{ \frac{u\lambda_1^{(p)}}{r} }^p }. \]

    For any non-zero $\vec y \in \cL$, we have that \[ f_p(\vec y, r) = p \paren{\frac{\lambda_1^{(p)}}{r}}^p \int_{\norm{\vec y}_p / \lambda_1^{(p)}}^{\infty} u^{p-1} g(u) du . \]
    
  Since $\norm{\vec y}_p/\lambdaonep\ge1$ for every nonzero $\vec y\in\cL$, summing over these $\vec y$ gives
    \begin{align*}
        f_p(\cL, r) &= 1+\sum_{\vec y \in \cL\setminus\{\vec 0\}} \exp\paren{- \frac{\norm{\vec y}_p^p}{r^p}} \\
        &= 1 +  p\left(\frac{\lambda_1^{(p)}}r\right)^p \int_1^\infty u^{p-1}N(u)g(u)\,du \\ 
        &\le 1+ p\left(\frac{\lambda_1^{(p)}}r\right)^p \int_1^\infty u^{p-1}(1+2u)^k g(u)\,du,
    \end{align*}
    where in the last inequality we have used \cref{eqn:packing}. Substitute \(v=u^p\), so \(dv=pu^{p-1}\,du\), in the above to obtain 
\[
f_p(\mathcal L,r)
\le
1+
\left(\frac{\lambda_1^{(p)}}r\right)^p
\int_1^\infty
(1+2v^{1/p})^k
e^{-(\lambda_1^{(p)}/r)^p v}\,dv.
\]Since \(p\ge1\) and \(v\ge1\), we have \(v^{1/p}\le v\). Thus
\[
f_p(\mathcal L,r)
\le
1+
\left(\frac{\lambda_1^{(p)}}r\right)^p
\int_1^\infty
(1+2v)^k
e^{-(\lambda_1^{(p)}/r)^p v}\,dv.
\]

Now we will bound the integral on the right. Substitute
\[
t=\frac{1}{2}\left(\frac{\lambda_1^{(p)}}r\right)^p(1+2v).
\]Then
\[
dv=\left(\frac r{\lambda_1^{(p)}}\right)^pdt,
\qquad
1+2v=2\left(\frac r{\lambda_1^{(p)}}\right)^pt,
\]and
\[
e^{-(\lambda_1^{(p)}/r)^p v}
=
e^{(\lambda_1^{(p)}/r)^p/2}e^{-t}.
\]Consequently,
\[
\begin{aligned}
f_p(\mathcal L,r)
&\le
1+
2^k\left(\frac r{\lambda_1^{(p)}}\right)^{pk}
e^{(\lambda_1^{(p)}/r)^p/2}
\int_{\frac32(\lambda_1^{(p)}/r)^p}^{\infty}
t^ke^{-t}\,dt\\
&\le
1+
2^k\left(\frac r{\lambda_1^{(p)}}\right)^{pk}
e^{(\lambda_1^{(p)}/r)^p/2}
\int_0^\infty t^ke^{-t}\,dt\\
&=
1+
2^k k!\left(\frac r{\lambda_1^{(p)}}\right)^{pk}
e^{(\lambda_1^{(p)}/r)^p/2}
\end{aligned}
\]
where the last integral equals \(k!\), by repeated integration by parts. 
\end{proof}

We prove \Cref{thm:main-svp} in the following form.

\begin{theorem}
    \label{thm:svp}
    Let $p \in [1, 2)$. There is an algorithm that, given a basis of a lattice $\cL \subseteq \Q^n$ of rank $k$, outputs a shortest nonzero vector of $\cL$ in the $\ell_p$ norm with probability at least $1-e^{-k}$.
    It runs in time $2^{\sigma k + o(k)}\poly(|I|)$, where $\sigma:=1+\log(2+1/\sqrt2)<2.4368$.
\end{theorem}

We apply \Cref{lem:upper_bound_fpl} at the following value of $r$.
 
\begin{corollary}
    \label{cor:fpl}
    Let $p\in[1,2)$ and $c>0$, let $\cL\subseteq\R^n$ be a lattice of rank $k$, let $\lambda:=\lambda_1^{(p)}(\cL)$, let $\lambda\le\lambda'\le(1+\frac1{k+2})\lambda$, and let $r:=\lambda'/(ck)^{1/p}$. Then $b:=(\lambda/r)^p$ satisfies $\frac{ck^2}{k+2}\le b\le ck$, and
    \[ f_p(\cL,r)\le1+e^3\sqrt k\,\Bigl(\frac2c\,e^{c/2-1}\Bigr)^k . \]
\end{corollary}
 
\begin{proof}
    We have $b=ck(\lambda/\lambda')^p$. Since $\lambda'\ge\lambda$, $b\le ck$. Since $p\le2$ and $(1+x)^{-2}\ge1-2x$, we have $(\lambda/\lambda')^p\ge(1+\frac1{k+2})^{-2}\ge\frac{k}{k+2}$, so $b\ge\frac{ck^2}{k+2}$. By \Cref{lem:upper_bound_fpl}, $f_p(\cL,r)-1\le2^k\,k!\,b^{-k}e^{b/2}$. Using $k!\le e\sqrt k\,(k/e)^k$, $b\ge\frac{ck^2}{k+2}$, $e^{b/2}\le e^{ck/2}$ and $(1+\frac2k)^k\le e^2$,
    \[ f_p(\cL,r)-1\le e\sqrt k\,\Bigl(1+\frac2k\Bigr)^k\Bigl(\frac{2}{ec}\Bigr)^ke^{ck/2}\le e^3\sqrt k\,\Bigl(\frac2c\,e^{c/2-1}\Bigr)^k . \qedhere \]
\end{proof}

\begin{algorithm}[H]
    \caption{$\svp_p$ for $p \in [1, 2)$}
    \label{alg:svp}
    \algorithmicrequire{A basis $B$ of a lattice $\cL\subseteq\Q^n$ of rank $k$.}
 
    \algorithmicensure{A shortest nonzero vector of $\cL$ in the $\ell_p$ norm.}
 
    \begin{algorithmic}[1]
        \State $c\gets\frac{2}{1+2\sqrt2}$, \quad $\gamma\gets1+\frac{1}{k+2}$, \quad $K\gets(n+2)^2$, \quad $N\gets\ceil{88\,k^{3/2}\,(2+1/\sqrt2)^k}$
        \State Compute an LLL-reduced basis of $\cL$, and let $\ell$ be the Euclidean length of its first vector
        \State $\cC\gets \set{\vec b_1}$, where $\vec b_1$ is the first vector in the LLL reduced basis. 
        \For{$j=0,1,\dots,K$}
            \State $\lambda'\gets2^{-(k-1)/2}\,\ell\,\gamma^j$, \quad $r\gets\lambda'/(ck)^{1/p}$
            \For{$i=1,\dots,N$}
                \State Sample $\vec s$ from the density $h_p(\cdot,r)$ \Comment{\Cref{lem:sample_hpr}}
                \State Run the algorithm of \Cref{thm:adrs_sampler} on $\cL_{\vec s}:=\diag(1/s_1,\ldots,1/s_n)\cL$ with parameter $1$ and $\kappa:=C_1k$; multiply each output vector by $\diag(s_1,\ldots,s_n)$, and add it to $\cC$ \label{itm:svp_alg_adrs_step}
            \EndFor
        \EndFor
        \State \Return a vector of least $\ell_p$ norm in $\cC\setminus\{\vec 0\}$
    \end{algorithmic}
\end{algorithm}
Here $C_1$ is a constant large enough that the statistical distance in \Cref{thm:adrs_sampler} is at most
\[ \varepsilon:=\frac{(2+1/\sqrt2)^{-k}}{88\sqrt k}\qquad\text{for every }k\ge1. \]
 
  \begin{proof}[Proof of \Cref{thm:svp}]
    Let $\vec v$ be a shortest nonzero vector of $\cL$ in the $\ell_p$ norm, and $\lambda:=\norm{\vec v}_p=\lambda_1^{(p)}(\cL)$.
 
    \textbf{The right guess.} For $p\in[1,2]$ we have $\norm{\cdot}_2\le\norm{\cdot}_p\le\sqrt n\,\norm{\cdot}_2$, and the first vector of an LLL-reduced basis has Euclidean length $\ell\le2^{(k-1)/2}\lambda_1^{(2)}(\cL)$ \cite{LLL82}. Hence $\lambda'_0:=2^{-(k-1)/2}\ell\le\lambda\le\sqrt n\,\ell$. Let $j$ be the least index such that $\lambda'_j:=\lambda'_0\gamma^j\ge\lambda$; then $\lambda\le\lambda'_j\le\gamma\lambda$. Moreover $j\le K$, because $\gamma^K\ge2^{(k-1)/2}\sqrt n\ge\lambda/\lambda'_0$; indeed, $K\ln\gamma\ge\frac{(n+2)^2}{k+3}\ge\frac{(n+2)^2}{n+3}\ge n+1\ge\ln(2^{(k-1)/2}\sqrt n)$. We consider the iterations of the outer loop with this $j$; the other iterations only add vectors of $\cL$ to $\cC$, which cannot hurt.
 
    \textbf{One trial.} Fix $\vec s$, and let $q(\vec s):=\rho(\vec v,\vec s)/\rho(\cL,\vec s)$. Since $\rho(\vec x,\vec s)=e^{-\pi\norm{\diag(1/s_1,\ldots,1/s_n)\vec x}^2}$, $q(\vec s)$ is the probability of $\diag(1/s_1,\ldots,1/s_n)\vec v$ under $D_{\cL_{\vec s},1}$. Let $m$ be the number of vectors output by the sampler; $m\ge\floor{2^{k/2}}\ge2^{k/2-1}$. If these were $m$ independent samples from $D_{\cL_{\vec s},1}$, then $\vec v$ would be missed with probability $(1-q)^m\le e^{-mq}\le1/(1+mq)$. Hence $\vec v$ is added to $\cC$ with probability at least
    \[ \frac{mq}{1+mq}-\varepsilon=\frac{\rho(\vec v,\vec s)}{\rho(\cL,\vec s)/m+\rho(\vec v,\vec s)}-\varepsilon . \]

  Now average over $\vec s\sim h_p(\cdot,r)$. For random variables $X\ge0$ and $Y>0$ we have $\expect[X^2/Y]\ge(\expect X)^2/\expect Y$, by the Cauchy--Schwarz inequality applied to $X= \frac{X}{\sqrt Y}\cdot\sqrt Y$. With $X=\rho(\vec v,\vec s)^{1/2}$, this gives
    \[ \Pr[\vec v\text{ is found in this trial}]\ge\frac{\paren{\expect\rho(\vec v,\vec s)^{1/2}}^2}{\expect\rho(\cL,\vec s)/m+\expect\rho(\vec v,\vec s)}-\varepsilon . \]
    By \cref{eqn:fp_as_marginal}, $\expect\rho(\vec x,\vec s)=f_p(\vec x,r)$ for every $\vec x\in\R^n$. Hence $\expect\rho(\cL,\vec s)=f_p(\cL,r)$, $\expect\rho(\vec v,\vec s)=f_p(\vec v,r)$, and $\expect\rho(\vec v,\vec s)^{1/2}=\expect\rho(\vec v/\sqrt2,\vec s)=f_p(\vec v/\sqrt2,r)$. Since $1/m\le2^{1-k/2}$,
    \begin{equation}
        \label{eq:svp_trial}
        \Pr[\vec v\text{ is found in this trial}]\ge\frac{f_p(\vec v/\sqrt2,r)^2}{2^{1-k/2}f_p(\cL,r)+f_p(\vec v,r)}-\varepsilon .
    \end{equation}

   \textbf{Numerator and denominator.} Let $b:=(\lambda/r)^p$; by \Cref{cor:fpl}, $\frac{ck^2}{k+2}\le b\le ck$. Then $f_p(\vec v,r)=e^{-b}\le1$, and, since $2^{1-p/2}\le\sqrt2$ for $p\ge1$, $f_p(\vec v/\sqrt2,r)^2=e^{-2^{1-p/2}b}\ge e^{-\sqrt2\,ck}$. The choice $c=\frac{2}{1+2\sqrt2}$ gives $c(\sqrt2+\frac12)=1$, that is, $\frac c2-1=-\sqrt2c$. So, by \Cref{cor:fpl},
    \[ 2^{1-k/2}f_p(\cL,r)\le2^{1-k/2}+2e^3\sqrt k\,\Lambda^k,\qquad\text{where }\Lambda:=\frac{\sqrt2}{c}\,e^{c/2-1}=\frac{\sqrt2}{c}\,e^{-\sqrt2c}\approx1.2932 . \]
    Since $\Lambda>1$, the denominator in \cref{eq:svp_trial} is at most $(2+2e^3+1)\sqrt k\,\Lambda^k\le44\sqrt k\,\Lambda^k$. Since $e^{-\sqrt2c}/\Lambda=c/\sqrt2=(2+1/\sqrt2)^{-1}$,
    \[ \Pr[\vec v\text{ is found in this trial}]\ge\frac{(2+1/\sqrt2)^{-k}}{44\sqrt k}-\varepsilon=\frac{(2+1/\sqrt2)^{-k}}{88\sqrt k}. \]

    \textbf{Conclusion.} For the right guess, the $N$ trials are independent, so $\vec v$ is missed by all of them with probability at most
    \[ \paren{1-\frac{(2+1/\sqrt2)^{-k}}{88\sqrt k}}^N\le\exp\paren{-\frac{N}{88\sqrt k\,(2+1/\sqrt2)^k}}\le e^{-k}. \]
    If $\vec v\in\cC$, the output is a shortest nonzero vector.

   \textbf{Running time.} LLL takes time $\poly(|I|)$. There are $(K+1)N$ trials. Each samples $\vec s$ in time $O(n)$ (\Cref{lem:sample_hpr}), calls the algorithm of \Cref{thm:adrs_sampler} with $\kappa=O(k)$, at cost $2^{k+o(k)}\poly(|I|)$, and computes $O(2^{k/2})$ vectors and their $\ell_p$ norms. Since $(K+1)N=O(n^2k^{3/2}(2+1/\sqrt2)^k)$, the total time is $2^{\sigma k+o(k)}\poly(|I|)$, where $\sigma=1+\log(2+1/\sqrt2)<2.4368$.
\end{proof}

\end{document}